\documentclass[9pt,conference]{IEEEtran}
\IEEEoverridecommandlockouts
\usepackage{longfigure}
\usepackage{fullpage}
\usepackage{enumitem}
\usepackage[bookmarks=false]{hyperref}
\usepackage{amsthm}
\usepackage[square, comma, sort&compress, numbers]{natbib}
\usepackage{amsmath,amssymb,amsfonts}
\usepackage{algorithmic}
\usepackage{graphicx}
\usepackage{textcomp}
\usepackage{subfigure}
\usepackage{float}
\usepackage[version=4]{mhchem}
\usepackage{xcolor}

\usepackage{xcolor}

    \makeatletter 
  \newcommand\figcaption{\def\@captype{figure}\caption} 
  \newcommand\tabcaption{\def\@captype{table}\caption} 
\makeatother
\begin{document}

\title{Molecular MUX-Based Physical Unclonable Functions\\
}


\author{\IEEEauthorblockN{Lulu Ge, and Keshab K. Parhi, {\em Fellow, IEEE}}
\IEEEauthorblockA{
Dept. of Electrical and Computer Engineering, University of Minnesota, Minneapolis, MN, USA \\
Email: \{ge000567, parhi\}@umn.edu}
}

\maketitle

\begin{abstract}
Physical unclonable functions (PUFs) are small circuits that are widely used as hardware security primitives for authentication. These circuits can generate unique signatures because of the inherent randomness in manufacturing and process variations. This paper introduces molecular PUFs based on multiplexer (MUX) PUFs using {\em dual-rail} representation. It may be noted that molecular PUFs have not been presented before. Each molecular multiplexer is synthesized using 16 molecular reactions. The intrinsic variations of the \textit{rate constants} of the molecular reactions are assumed to provide {\em inherent} randomness necessary for uniqueness of PUFs. Based on Gaussian distribution of the rate constants of the reactions, this paper simulates intra-chip and inter-chip variations of linear molecular MUX PUFs containing $8$, $16$, $32$ and $64$ stages. These variations are, respectively, used to compute reliability and uniqueness. It is shown that, for the rate constants used in this paper, although $8$-state molecular MUX PUFs are not useful as PUFs, PUFs containing 16 or higher stages are useful as molecular PUFs. Like electronic PUFs, increasing the number of stages increases uniqueness and reliability of the PUFs.
\end{abstract}

 \begin{IEEEkeywords}
Physical unclonable function (PUF), Molecular PUF, Molecular MUX PUF, Molecular multiplexers, dual-rail encoding, biomolecular security.
\end{IEEEkeywords}
\vspace{-8pt}
\section{Introduction}\label{s:1}
Physical unclonable functions (PUFs) have attracted increased attention in both academia and industry for their unique unclonability---these are easy to be created but difficult to be reproduced due to their intrinsic manufacturing process variations \cite{pappu2002physical,gassend2002silicon,suh2007physical,lao2014statistical,gao2016emerging}. These variations are sufficient for each PUF to generate unique signatures. For this reason, PUFs are widely used as hardware security primitives \cite{arppe2017physical}. 

There are a variety of PUFs \cite{armknecht2011formalization}, which can be coarsely classified based on whether the randomness is introduced extrinsically or intrinsically. Extrinsic-based PUFs include optical PUF and coating PUF, while delay PUF, radio frequency PUF \cite{chatterjee2018rf} and magnetic PUF are typical intrinsic-based PUFs. Among various PUF implementations, multiplexer-based PUF (MUX-based PUF), a type of delay PUFs, is of great interest as it can be easily constructed and analyzed. 

Molecular computing involves computations via molecules, such as DNA, rather than silicon substrate for electrical computers. As a programming language for molecular computing \cite{johnson2016verifying}, chemical reaction networks (CRNs) specify a set of chemical species and a set of specific chemical reactions. The deduced system behavior is obtained by the CRN model via ordinary differential equations (ODEs) \cite{cardelli2008processes}. Such a CRN model can be applied to real molecular system analysis \cite{soloveichik2010dna}, and also can simplify the process of designing engineered systems \cite{chen2013programmable, ge2017formal}. Guaranteed by \cite{soloveichik2010dna}, the designed formal CRN can be mapped to real DNA strand displacement reactions using unimolecular and bimolecular reactions. Similar to electronics, approaches to digital computing using molecular logic have been well established; examples include computing using combinational logic \cite{jiang2013digital,gao2016emerging}, sequential logic \cite{jiang2011synchronous,salehi2015molecular} and clock signal \cite{kharam2011binary}.

For anti-counterfeiting, interesting works for PUF implementation through chemical methods have been presented in \cite{arppe2017physical}. For example, the molecular tags, encrypted by nucleic acid's sequence, color and length, weaken the unclonability. Molecular PUFs can be used in numerous bio-security applications for authentication. One application might be authentication of oligonucleotide Arrays \cite{holden2019encrypted}. It is well known that a counterfeiter can easily decipher the array sequences. DNA barcoding has been used for preventing seafood fraud by species substitution \cite{shokralla2015dna}. With decreasing cost of DNA, the DNA PUFs may find applications in authentication of many chemical and biological substances where using electronic PUFs would be prohibitively expensive.

Although various types of PUFs have been presented in the literature, no molecular PUF has been presented so far. This paper introduces an approach to implement an $N$-stage molecular MUX-based PUF where the intrinsic randomness is derived from the {\em change in the rate constant} of the molecular reaction. All molecules in the MUX PUF are represented by {\em dual-rail} encoding \cite{qian2011scaling}. The paper introduces a proof of concept by simulations assuming the rate constant to vary in a Gaussian manner. This proof of concept describes feasibility of molecular PUF as a molecular security primitive. However, a practical demonstration requires experiment in a test tube. To be more specific, the rate constant of each single MUX follows the Gaussian distribution, i.e., $rate \sim \mathcal{N}(\mu, \sigma^2)$, which is assumed to be $\mathcal{N}(16, 1)$ throughout the entire paper. Using other rate constants is likely to lead to different conclusions. After constructing each single MUX with CRN using the one-to-one mapping method proposed in \cite{ge2017formal}, the entire PUF is synthesized by cascading all the multiplexers according to the designed PUF configuration. Four MUX PUFs, $8$, $16$, $32$, $64$-stage molecular PUFs, are simulated using molecular reactions, and their uniqueness and reliability properties are investigated. A single PUF should be able to generate different output responses $R$ under different challenges, $C$, while different PUFs should generate different $R$ for the same $C$. In addition, two PUF metrics, $Reliability$ and $Uniqueness$, are calculated. For a synthesized molecular PUF to be feasible, the design must satisfy the fact that the minimum inter-chip variation is larger than the maximum intra-chip variation. The impact of the number of MUX stages on the performance of the molecular PUF is also investigated. Therefore, this paper extends the PUF concept towards molecular computing with the CRN model, whose physical implementation could be DNA strand displacement reactions.

The remainder of this paper is organized as follows. Section \ref{s:2} briefly introduces the preliminaries. Two important metrics, $Reliability$ and $Uniqueness$, are also introduced in this section. Starting with a single MUX synthesis, Section \ref{s:3} illustrates how to implement an $N$-stage molecular PUF by introducing randomness in the rate constant of each single MUX CRN. \textcolor{black}{Totally four cases are studied; these include $8$, $16$, $32$ and $64$-stage PUFs.} Both reliability and uniqueness metrics, are also calculated in this section. The complexity is summarized in Section \ref{s:4}. Finally, Section \ref{s:5} concludes the entire paper.
\vspace{-6pt}

\section{Preliminaries}\label{s:2}
\vspace{-4pt}
\subsection{Chemical Reaction Networks (CRNs)}

Formal CRNs can be realized by DNA strand displacement reactions as long as the number of reactants per equation is no more than two \cite{soloveichik2010dna}. This paper, however, presents molecular PUFs; these have {\em not} been mapped to DNA. The chemical kinetics of the CRNs can be obtained by numerical simulation of ODEs based on mass-action law.

\subsection{Dual-Rail Representation}
Dual-rail representation refers to two species, e.g., $X_0$ and $X_1$, are employed to represent a single bit $X$ \cite{qian2011scaling}. If $X$ is $1$, then $X_1=~1$ and $X_0~=~0$. If $X$ is $0$, then $X_1=~0$ and $X_0~=~1$. All molecules used in this paper are encoded using dual-rail representation.
\subsection{MUX-based PUFs: Uniqueness and Reliability}
This paper studies the molecular MUX-based PUFs described by a linear delay model. The reliability and uniqueness properties of these PUFs are addressed.

\subsubsection{\textbf{The Structure of PUF}}
\begin{figure}[H]
\centerline{\includegraphics[width=0.99\linewidth]{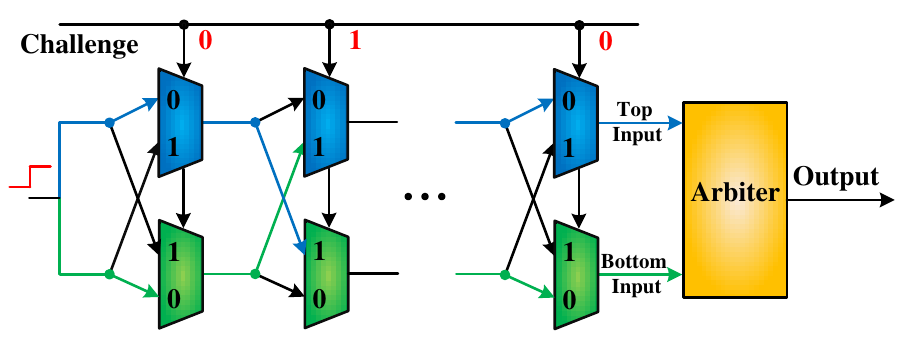}}
\vspace{-4pt}
\caption{Configuration of an $N$-stage MUX-based PUF.}
\label{f0}
\end{figure}

Fig. \ref{f0} shows the basic configuration of an $N$-stage MUX-based PUF. Given the $N$-bit challenge, then the clock signal races through both top path and bottom path of the cascaded $N$-stage multiplexers. The arbiter produces the output response $R$ according to whether the clock reaches the top or bottom input of the arbiter first. If the top input is activated first, then the output response $R=1$, otherwise $R=0$. 
\vspace{-8pt}
\subsubsection{\textbf{The Additive Linear Delay Model}}
\vspace{-4pt}

For each single MUX, the time delay can be defined as an independent identically distributed (i.i.d.) random variable $D_i$, which follows the Gaussian distribution $\mathcal{N}(\mu,\sigma^2)$, where $\mu$ is the mean and $\sigma$ is the standard deviation. At the $i^{th}$ stage, the delay difference $\Delta_i$ between top and bottom paths can be described by \eqref{eq0:0}. The delay difference $r_{N}$ after the last stage is modeled by \eqref{eq0:1}, where $C'_i=\oplus^{N}_{j=i+1} C_j$ and $C'_N=0$. The output response $R$ is computed by \eqref{eq0:2}. Interested readers can refer to \cite{lao2014statistical} for more details of such a linear delay model.
\begin{subequations}\label{eq0}
\begin{equation}\label{eq0:0}
\Delta_i=D^t_{i}-D^b_{i} \sim \mathcal{N}(0,2\sigma^2),
\end{equation}\vspace{-10pt}
\begin{equation}\label{eq0:1}
r_{N}=\sum_{i=1}^{N}(-1)^{C'_i}\Delta_i,
\end{equation}\vspace{-8pt}
\begin{equation}\label{eq0:2}
R=sign(r_{N})=
\begin{cases}
1,      &  r_{N} \geq 0, \\
0,      &  r_{N} < 0.
\end{cases}
\end{equation}
\end{subequations}

One thing should be emphasized is that a single $N$-bit challenge can produce a $1$-bit response $R$. Thus, $L$ challenges with $N$ bits are required to generate a signature of length $L$.

Note that all the multiplexers are designed identically. The main reason for the clock signal not arriving at the arbiter simultaneously is manufacturing process variations. Although all the multiplexers are designed to be identical, their rate constants affect the propagation delay. This means in practice we can never produce the same PUF. Therefore, even for the same challenge, different PUFs with the identical configuration will generate different output responses, also referred as signatures. Moreover, a PUF may generate different responses to the same challenge under various environmental and noise conditions.

\subsubsection{\textbf{Intra-Chip and Inter-Chip Variations}}
In general, environmental and noise conditions affect the intra-chip variation and the manufacturing process variations lead to inter-chip variation.

\paragraph{Intra-chip variation} This variation refers to variability for a single PUF. This intra-chip variation means that, for a certain PUF, if the challenge bits are fixed, the output response $R$ can vary under different environmental conditions. 

\paragraph{Inter-chip variation} This corresponds to the variation from chip to chip, i.e., PUF to PUF. This inter-chip variation implies that the signatures of different PUFs will be different for identical challenge. 

\subsubsection{\textbf{Reliability and Uniqueness}}
In this paper, two PUF metrics, namely reliability and uniqueness, are studied to quantify the PUF performance.




\paragraph{Reliability}

This metric measures the reliability of a single PUF when generating response bits under different environmental conditions \cite{maiti2013systematic}. The reliability metric is computed by \eqref{eq1}, where $P_{intra}$ reflects the intra-chip variation for the entire $L$-bit response. $P_{intra}$ is computed by \eqref{eq1:1} as the average Hamming distance (HD) between $L$-bit responses generated by the same PUF under $m$ different environmental conditions. The closer the value of $Reliability$ to $1$, the greater reliability for a PUF. 
\begin{subequations}\label{eq1:0}
\begin{equation}\label{eq1}
Reliability=1-P_{intra} , \quad P_{intra} \in [0,1],
\end{equation}\vspace{-8pt}
\begin{equation}\label{eq1:1}
\begin{aligned}
P_{intra}&=\mathbb{E} [HD_{intra}]\\
&=\mathbb{E}\left[ \frac{1}{m}\sum_{i=2}^{m}\dfrac{HD(R_1,R_i)}{L} \times 100\% \right].
\end{aligned}
\end{equation}
\end{subequations}
\paragraph{Uniqueness}
Computed by \eqref{eq2}, this metric quantifies the ability of a PUF to be uniquely distinguished from a group of PUFs with the same configuration \cite{maiti2013systematic}, where $P_{inter}$ reflects the inter-chip variation. Assume given $K$ PUF instances, $P_{inter}$ is computed using \eqref{eq2:2} as the average Hamming distance of all ${K \choose 2}$ possible pairs comparing combinations of $L$-bit responses. The better $Uniqueness$, the value is closer to $1$. $P_{inter}=50 \% $ represents the best uniqueness for a PUF.
\vspace{-10pt}
\begin{subequations}\label{eq2:0}
\begin{equation}\label{eq2}
\resizebox{.88\hsize}{!}{$
Uniqueness=1-\left| 2P_{inter}-1 \right|, \quad P_{inter} \in [0,1],
$}
\end{equation}\vspace{-8pt}
\begin{equation}\label{eq2:2}
\resizebox{.88\hsize}{!}{$
\begin{aligned}
P_{inter}&=\mathbb{E} [HD_{inter}]\\
&=\mathbb{E} \left[ \dfrac{2}{(K-1)K}\sum_{i=1}^{K-1}\sum_{j=i+1}^{K}\dfrac{HD(R_i,R_j)}{L} \times 100\% \right].
\end{aligned}
$}
\end{equation}
\end{subequations}

This paper only covers the reliability and uniqueness metrics for MUX-based PUF. Other types of PUF configuration like feed-forward MUX-based PUFs \cite{lao2014statistical} and other metrics like randomness, unpredictability, and security are also of interest, but these are not investigated in this paper.


\section{Molecular MUX-based PUF with Dual-Rail Representation}\label{s:3}


This section first presents the CRN implementation for a 2-to-1 multiplexer, the unit module for a PUF. Then the molecular PUF synthesis is illustrated in detail. For various PUF instances, both reliability and uniqueness are calculated to analyze the molecular PUF performance.

\subsection{CRN for a 2-to-1 MUX}
\begin{figure}[H]
\centerline{\includegraphics[width=0.8\linewidth]{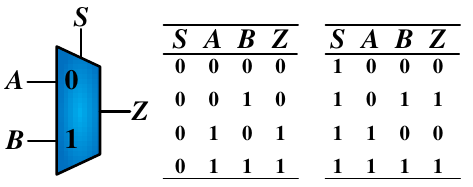}}
\caption{The truth table for a 2-to-1 multiplexer.}
\label{f1}
\end{figure}

The logic function of a 2-to-1 multiplexer is expressed by $Z=A \cdot \bar{S}+ B \cdot S$, where $A$ and $B$ are the two inputs, $S$ represents the select signal and $Z$ represents the output, where $\bar{S}$ means the \textsc{NOT} operation of $S$. The corresponding Truth Table is shown in Fig. \ref{f1}.
\vspace{-6pt}
\vspace{-4pt}
A 2-to-1 MUX can be synthesized using dual-rail encoding with one-to-one \cite{ge2017formal} mapping from the Truth Table to molecular reactions. Each molecular MUX is described by $16$ reactions listed in \eqref{eq3}, where $rate$ is a positive constant. Note chemical species $R_1$, $R_2$, $R_3$ and $R_4$ are intermediate variables. These are generated to make sure the number of reactants is no more than two. While bistable reactions \cite{jiang2013digital,ge2017formal} have been used in digital logic, we argue that bistable reactions should not be used for molecular PUFs. This is because the bistable reactions prevent racing of the input signal as the output of every stage is first guaranteed to be stable before propagating to next stage in these reactions. Before each authentication, the output $Z$ is set to logic $0$, i.e., the initial concentration for $Z_0$ is set as $100~nM$ and for $Z_1$ is set as $0~nM$.
\begin{subequations}\label{eq3}
\begin{equation}\label{eq3:0}
\left\{
 \begin{aligned}
  A_0 + B_0 & \ce{<=>[rate][10^5]}    R_1,\\ 
  A_0+B_1  &  \ce{<=>[rate][10^5]}   R_2,\\
\end{aligned}
\right.
\quad
\left\{
 \begin{aligned}
  A_1+B_0 & \ce{<=>[rate][10^5]}    R_3,\\ 
  A_1+B_1 & \ce{<=>[rate][10^5]}    R_4,\\
\end{aligned}
\right.
\end{equation}
\vspace{-2pt}
\begin{equation}\label{eq3:1}
\left\{
 \begin{aligned}
  S_0+R_1 & \stackrel{rate}{\longrightarrow} S_0+A_0 + B_0+Z'_0,\\ 
  S_0+R_2 & \stackrel{rate}{\longrightarrow} S_0+A_0+B_1+Z'_0,\\
  S_0+R_3 & \stackrel{rate}{\longrightarrow} S_0+A_1+B_0+Z'_1,\\
  S_0+R_4 & \stackrel{rate}{\longrightarrow} S_0+A_1+B_1+Z'_1,\\
   S_1+R_1 & \stackrel{rate}{\longrightarrow} S_1+A_0 + B_0 + Z'_0,\\ 
  S_1+R_2 & \stackrel{rate}{\longrightarrow} S_1+A_0+ B_1 + Z'_1,\\
  S_1+R_3 & \stackrel{rate}{\longrightarrow} S_1+A_1+B_0+Z'_0,\\
  S_1+R_4 & \stackrel{rate}{\longrightarrow} S_1+A_1+B_1+Z'_1,\\
\end{aligned}
\right.
\end{equation}
\vspace{-2pt}
\begin{equation}\label{eq3:2}
\left\{
 \begin{aligned}
 Z'_0+Z_1 \stackrel{rate}{\longrightarrow} Z_0,\\
 Z'_0+Z_0 \stackrel{rate}{\longrightarrow} Z_0,\\
\end{aligned}
\right.
\quad
\left\{
 \begin{aligned}
 Z'_1+Z_1 \stackrel{rate}{\longrightarrow} Z_1,\\
 Z'_1+Z_0 \stackrel{rate}{\longrightarrow} Z_1,\\
\end{aligned}
\right.
\end{equation}
\end{subequations}
\vspace{-2pt}



 
\subsection{CRN for a PUF}\label{sec:puf}
Based on the configuration shown in Fig. \ref{f0}, the molecular PUF is synthesized via constructing its all unit modules---multiplexers with the formal CRN as described by \eqref{eq3}. Manufacturing process variation is introduced by changing rate constants for each MUX in the PUF. The rate constants for the two MUXes in each stage and for different stages are different. This means a slight variation in rate constant is assumed to be the source of randomness for each stage. Similar to the delay difference $D_i$, the rate constant of each MUX, $rate$, follows a Gaussian distribution $\mathcal{N}(\mu, \sigma^2)$. Specifically, all the rate constants in this paper are randomly sampled from the probability density function $\mathcal{N}(16, 1)$. The response $R$ depends on the top and bottom outputs of the last MUX stage. Thus we only consider the $N^{th}$ stage outputs of the top and bottom paths, $Z^t_{1,N}$ and $Z^b_{1,N}$. If the input signal propagates to $Z^t_{1,N}$ first, then $R$ is $1$; otherwise it is $0$. Since we adopt dual-rail representation, two molecules are employed to represent a single bit. Considering the logic value is either $0$ or $1$, then the corresponding initial species concentration is ideally $0$ or $100~nM$. The clock signal shown in Fig. \ref{f0} reaches the top and bottom inputs of the arbiter but with a delay difference. 

In this paper, a molecular arbiter has not been implemented as part of the molecular PUF. Instead, the response is computed as $R ~ = ~ Z^t_{1,N} - Z^b_{1,N}$ in software. 




\subsubsection{Intra-Chip Variation of a PUF under Different Environmental Conditions}

\begin{figure}[H]
  \centering
  \subfigure[\textcolor{black}{8-stage PUF (Cond. 1) (R=1)}.]{
 \label{f3_1}
 \includegraphics[width=0.475\linewidth]{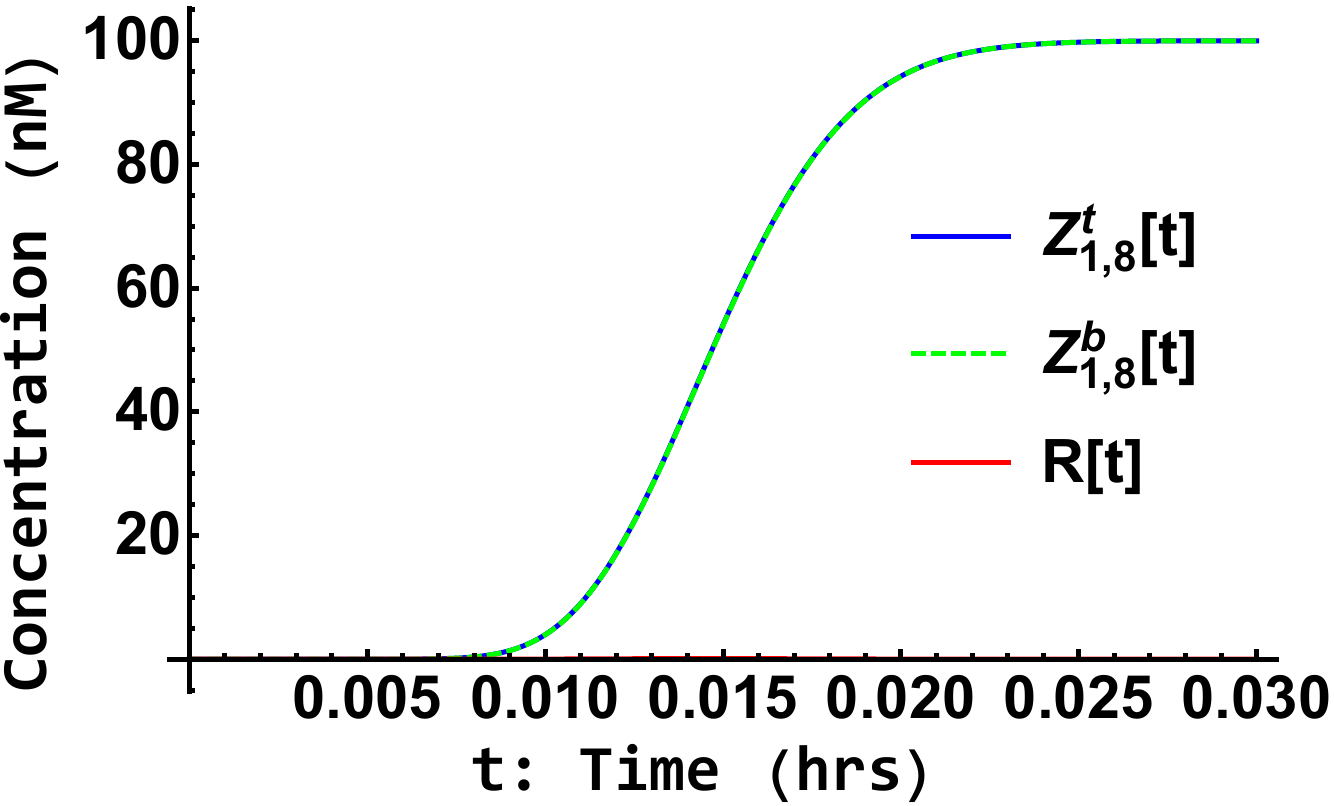}}
  \vspace{-4pt}
  \subfigure[\textcolor{black}{Enlarged version of Cond. 1 (R=1).}]{
 \label{f3_2}
 \includegraphics[width=0.475\linewidth]{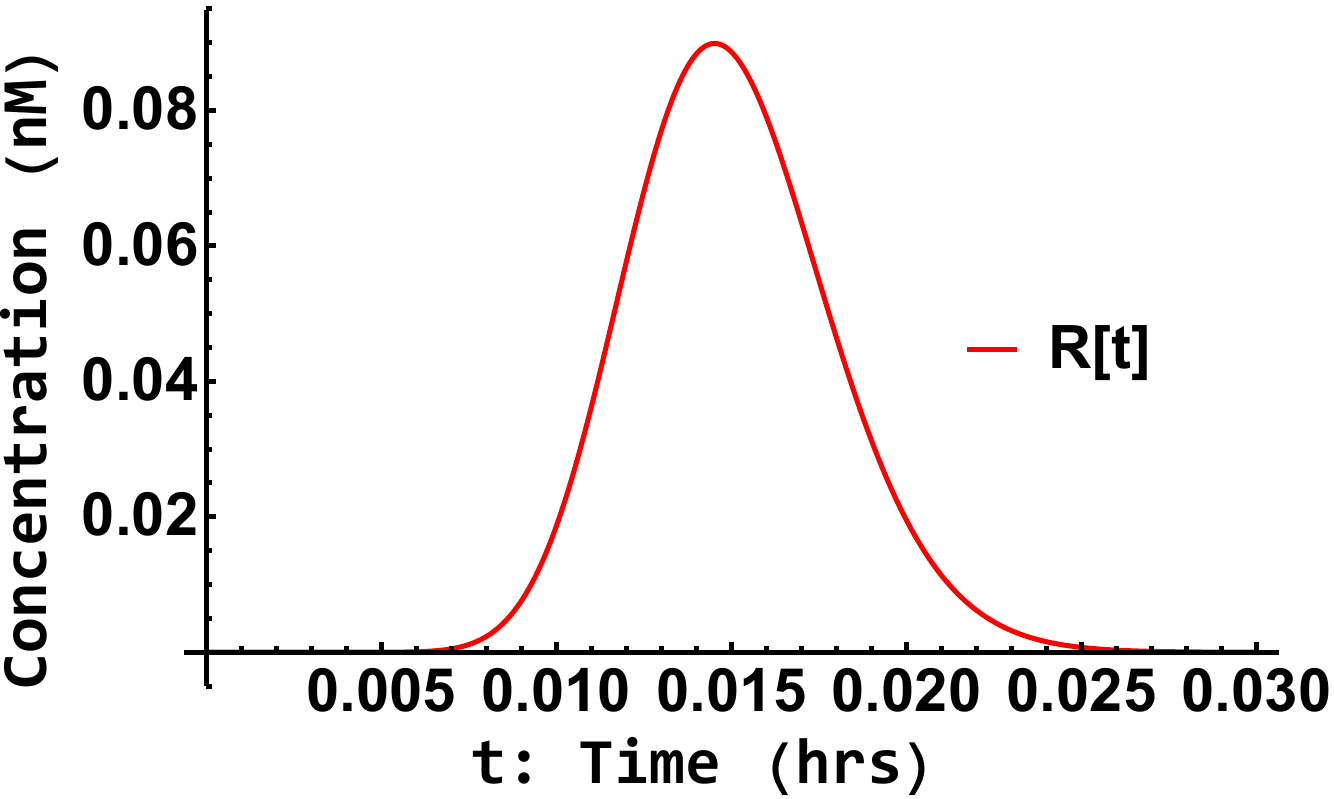}}
  \vspace{-4pt}
 \subfigure[8-stage PUF (Cond. 2)  (R=0).]{
 \label{f3_3}
  \includegraphics[width=0.475\linewidth]{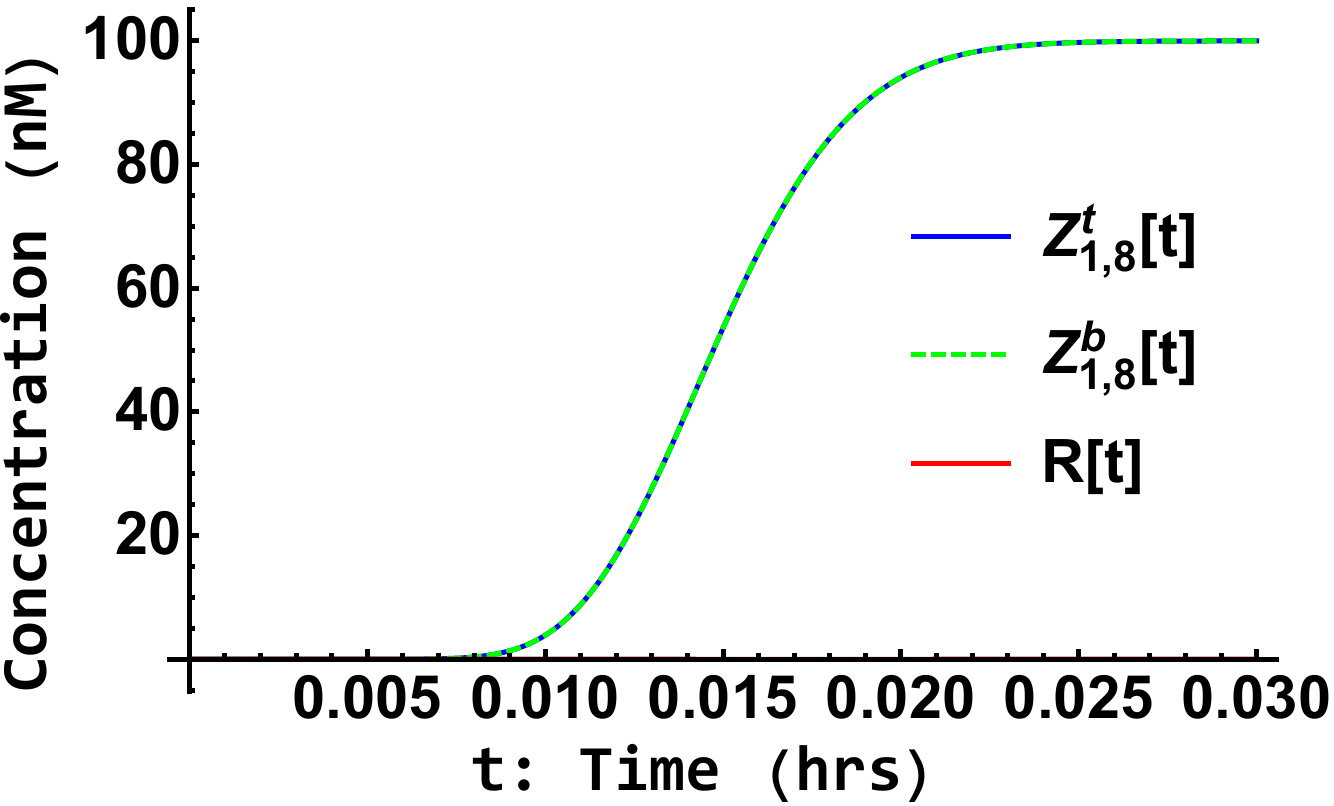}}
   \vspace{-4pt}
  \subfigure[Enlarged version of Cond. 2 (R=0).]{
 \label{f3_4}
  \includegraphics[width=0.475\linewidth]{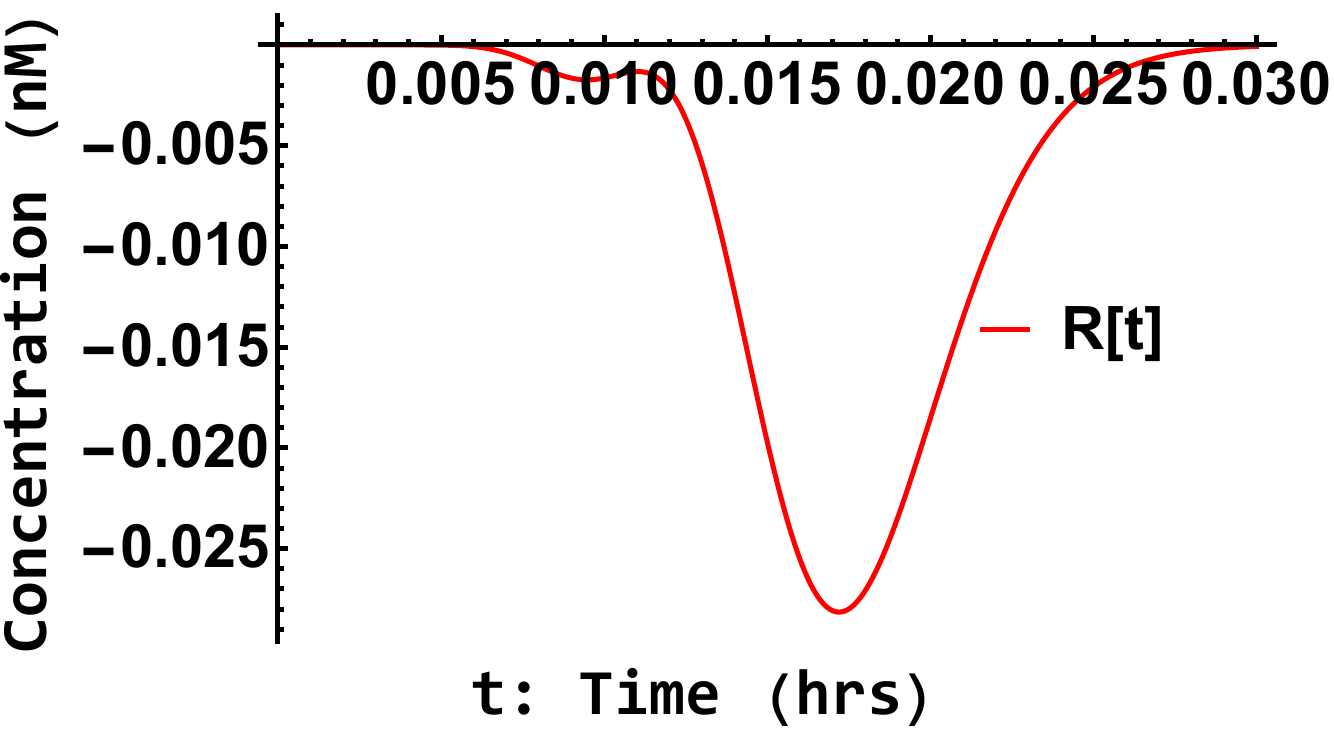}}
   \vspace{-4pt}
 \subfigure[ 16-stage PUF (Cond. 1) (R=1).]{
 \label{f3_5}
 \includegraphics[width=0.475\linewidth]{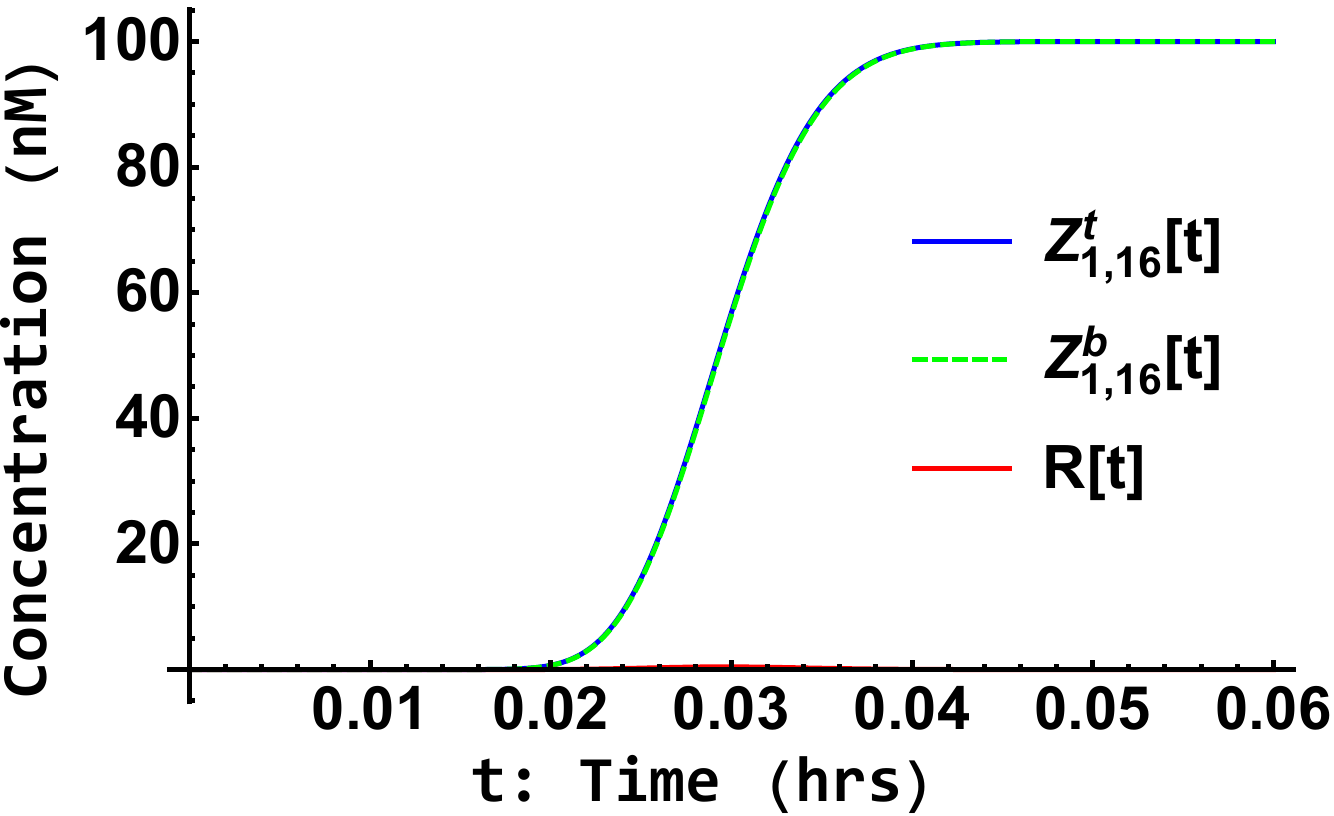}}
  \vspace{-4pt}
  \subfigure[\textcolor{black}{Enlarged version of Cond. 1  (R=1)}.]{
 \label{f3_6}
 \includegraphics[width=0.475\linewidth]{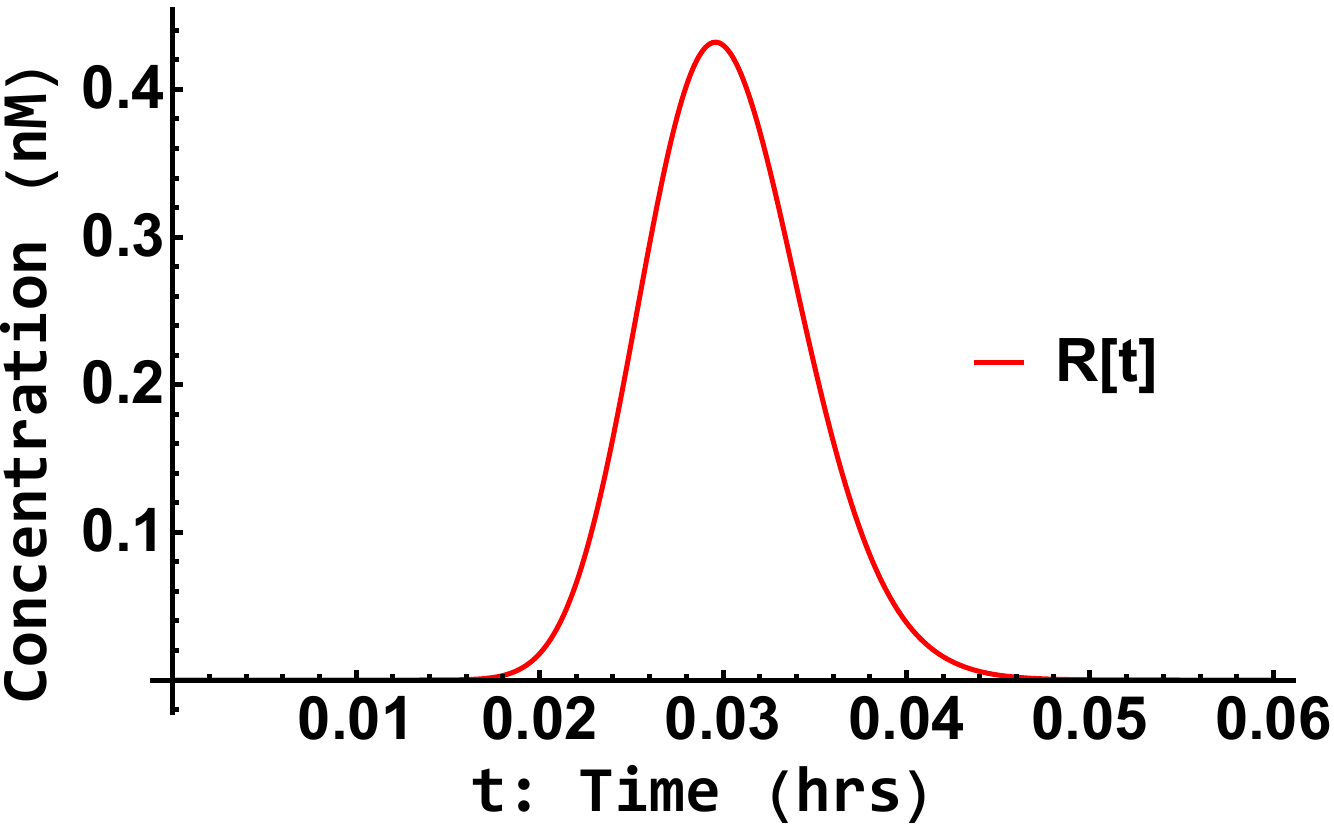}}
 \vspace{-4pt}
 \subfigure[16-stage PUF (Cond. 2)  (R=0).]{
 \label{f3_7}
  \includegraphics[width=0.475\linewidth]{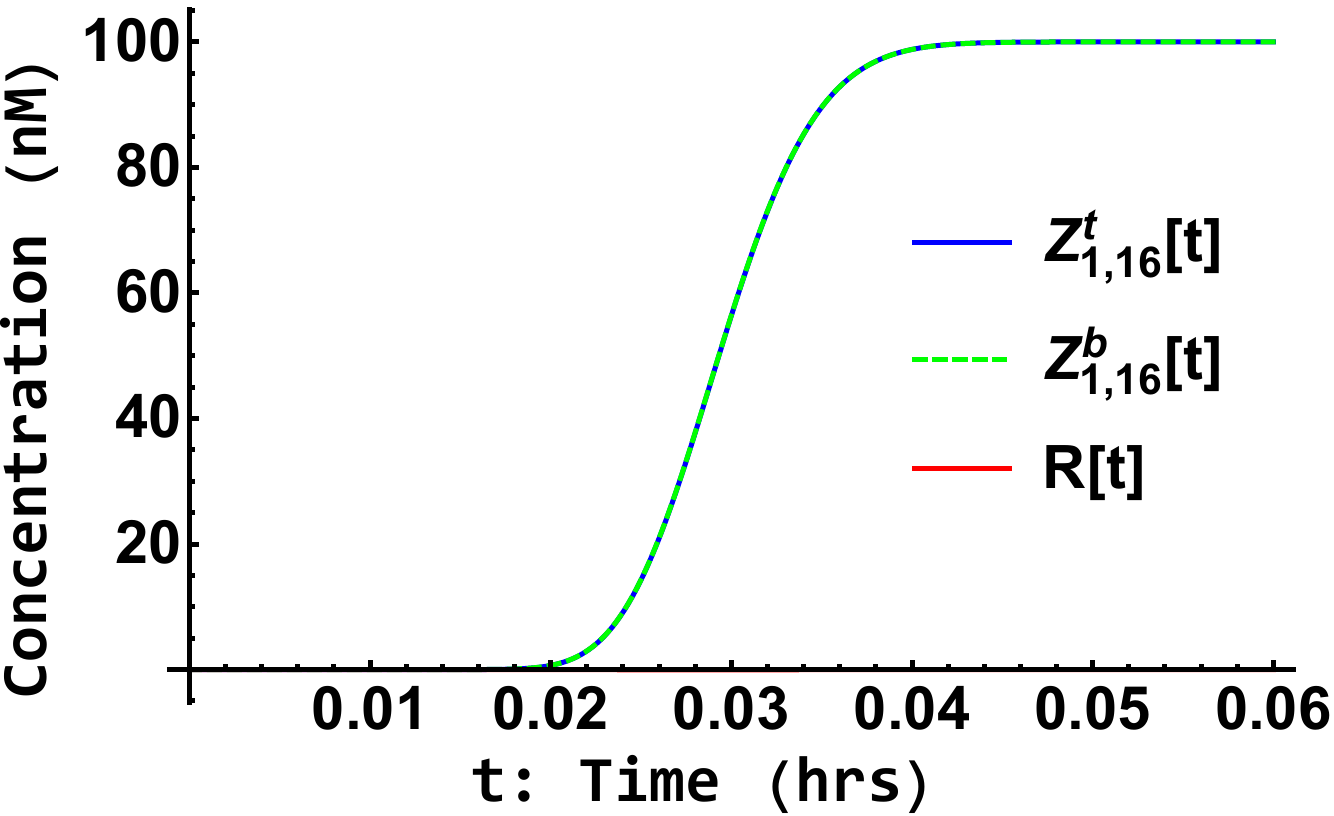}}
     \vspace{-4pt}
  \subfigure[Enlarged version of Cond. 2  (R=0).]{
 \label{f3_8}
  \includegraphics[width=0.475\linewidth]{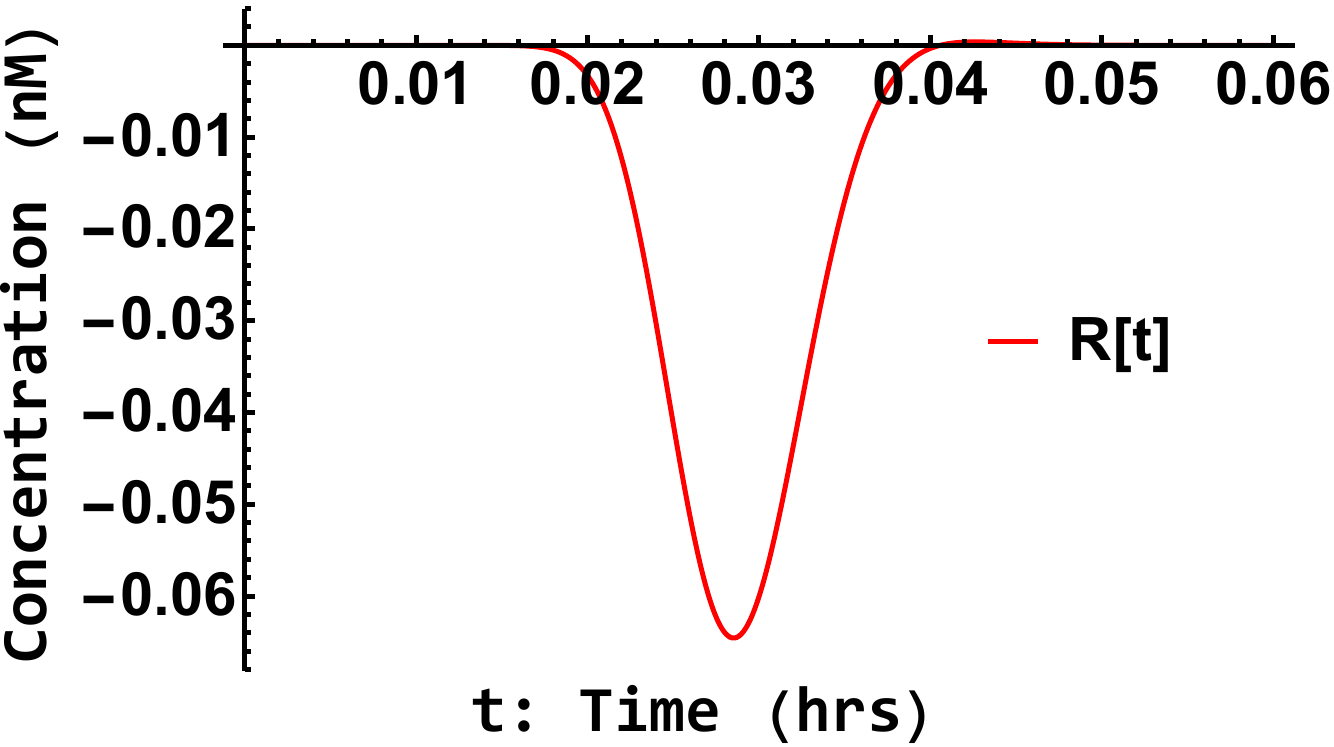}}
  \vspace{-2pt}
 \caption{Simulations for a single $8$, $16$, $32$ and $64$-stage PUF driven by the same challenge under two different environmental conditions denoted as Cond. 1 and Cond. 2.}\label{f3}
\end{figure}

Fig. \ref{f3} shows four cases where a single PUF is activated by the given challenge under two different environmental conditions, which produces two different responses. The simulation results show our synthesized molecular PUF has the ability to produce different responses under various conditions even when driven by the same challenge.
\vspace{-12pt}
\addtocounter{figure}{-1}       
\begin{figure}
  \centering 
  \addtocounter{subfigure}{2} 
\subfigure[\textcolor{black}{32-stage PUF (Cond.1) (R=1)}.]{
 \label{f3_9}
 \includegraphics[width=0.475\linewidth]{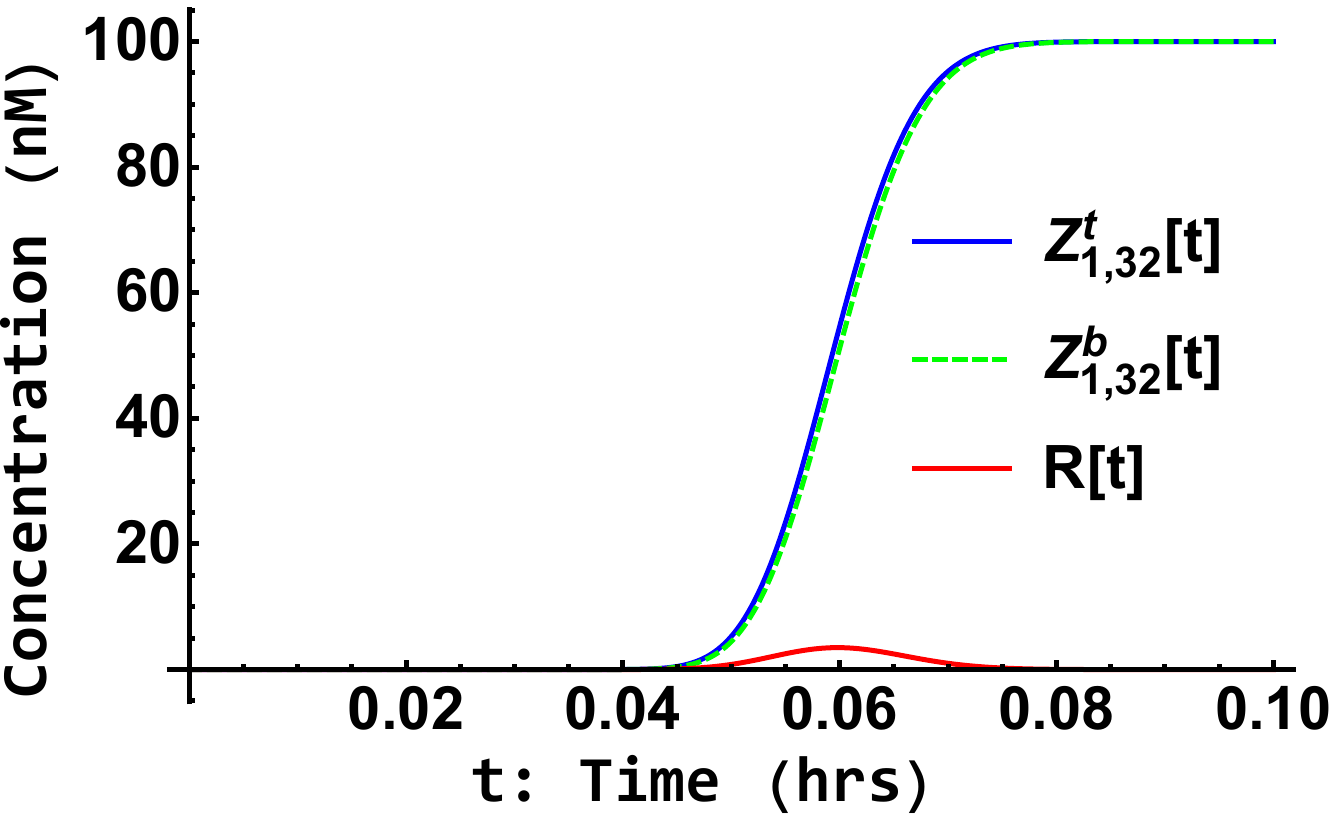}}
    \vspace{-4pt}
  \subfigure[\textcolor{black}{Enlarged version of Cond. 1 (R=1)}.]{
 \label{f3_10}
 \includegraphics[width=0.475\linewidth]{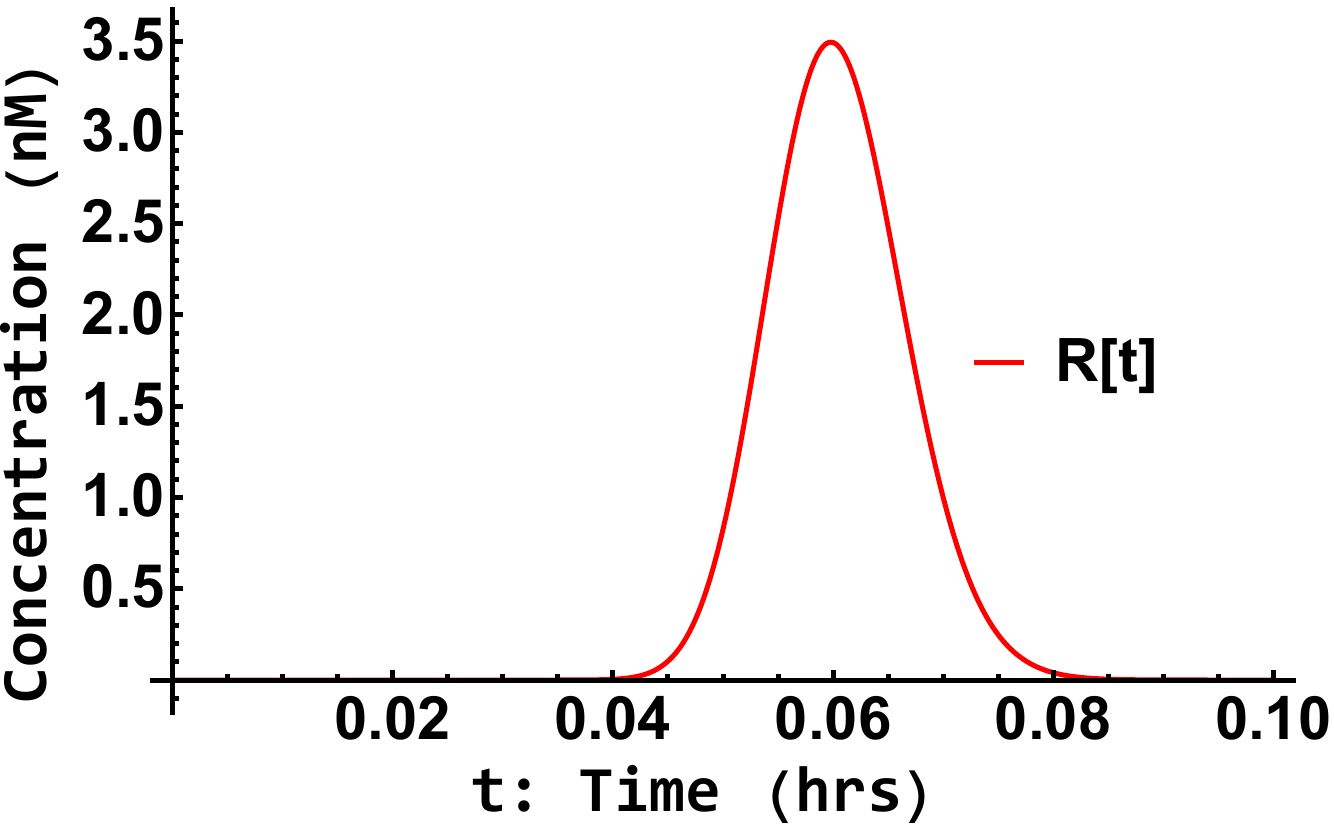}}
    \vspace{-4pt}
 \subfigure[32-stage PUF (Cond.2) (R=0).]{
 \label{f3_11}
  \includegraphics[width=0.475\linewidth]{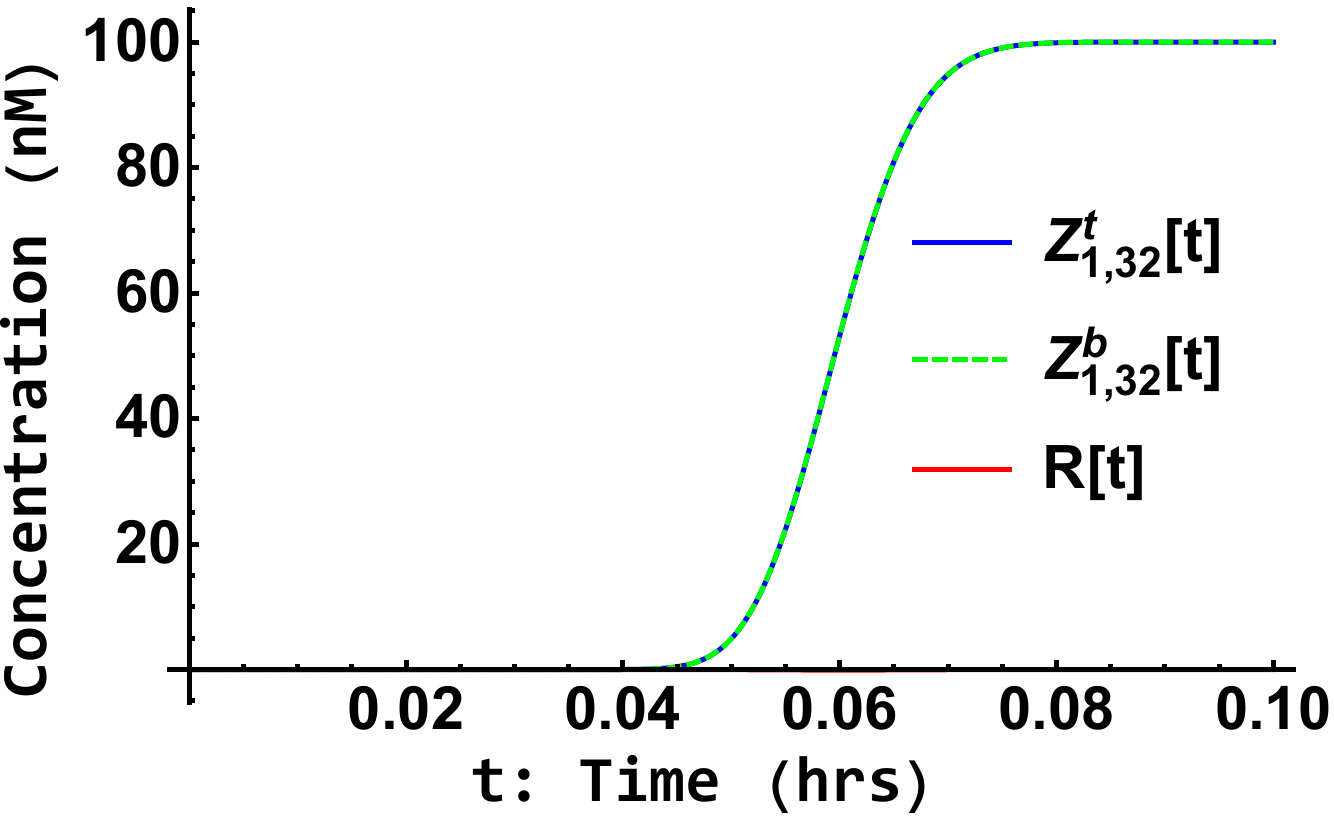}}
     \vspace{-4pt}
  \subfigure[Enlarged version of Cond. 2 (R=0).]{
 \label{f3_12}
  \includegraphics[width=0.475\linewidth]{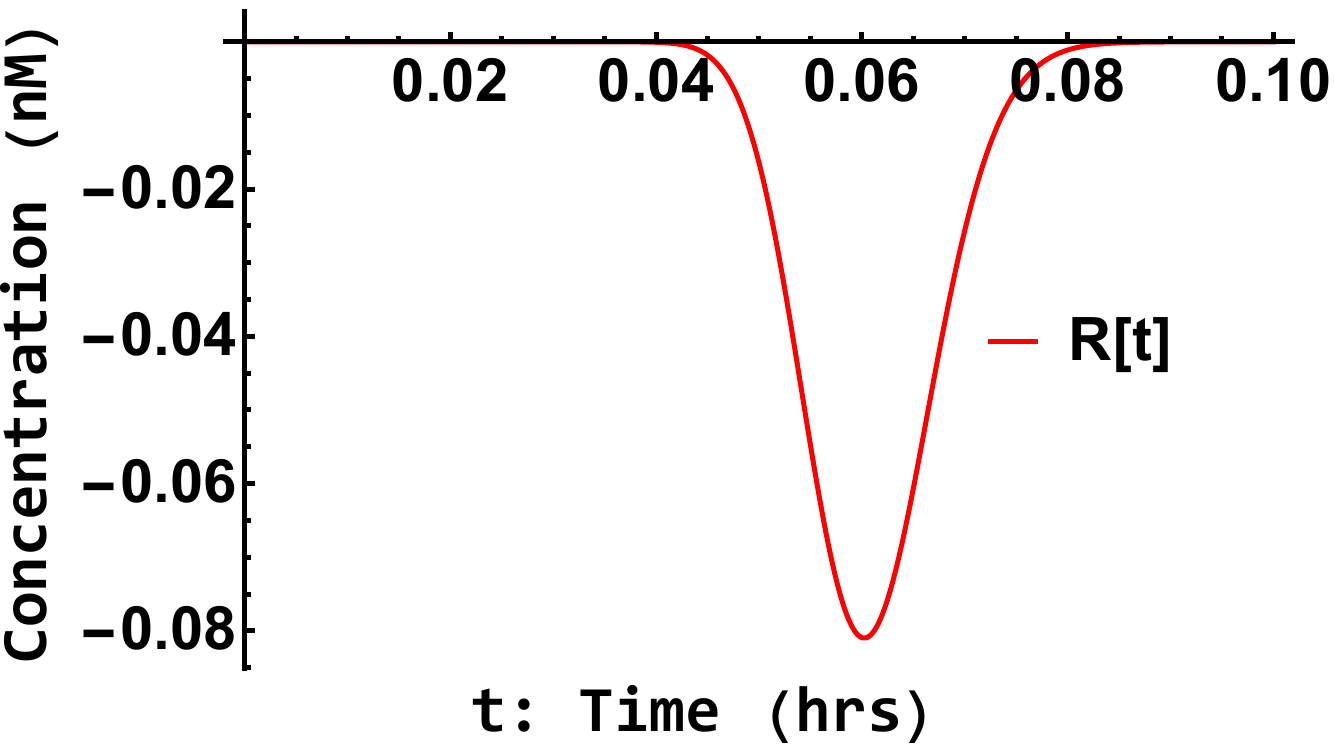}}
     \vspace{-4pt}
\subfigure[\textcolor{black}{ 64-stage PUF (Cond.1) (R=1)}.]{
 \label{f3_13}
 \includegraphics[width=0.475\linewidth]{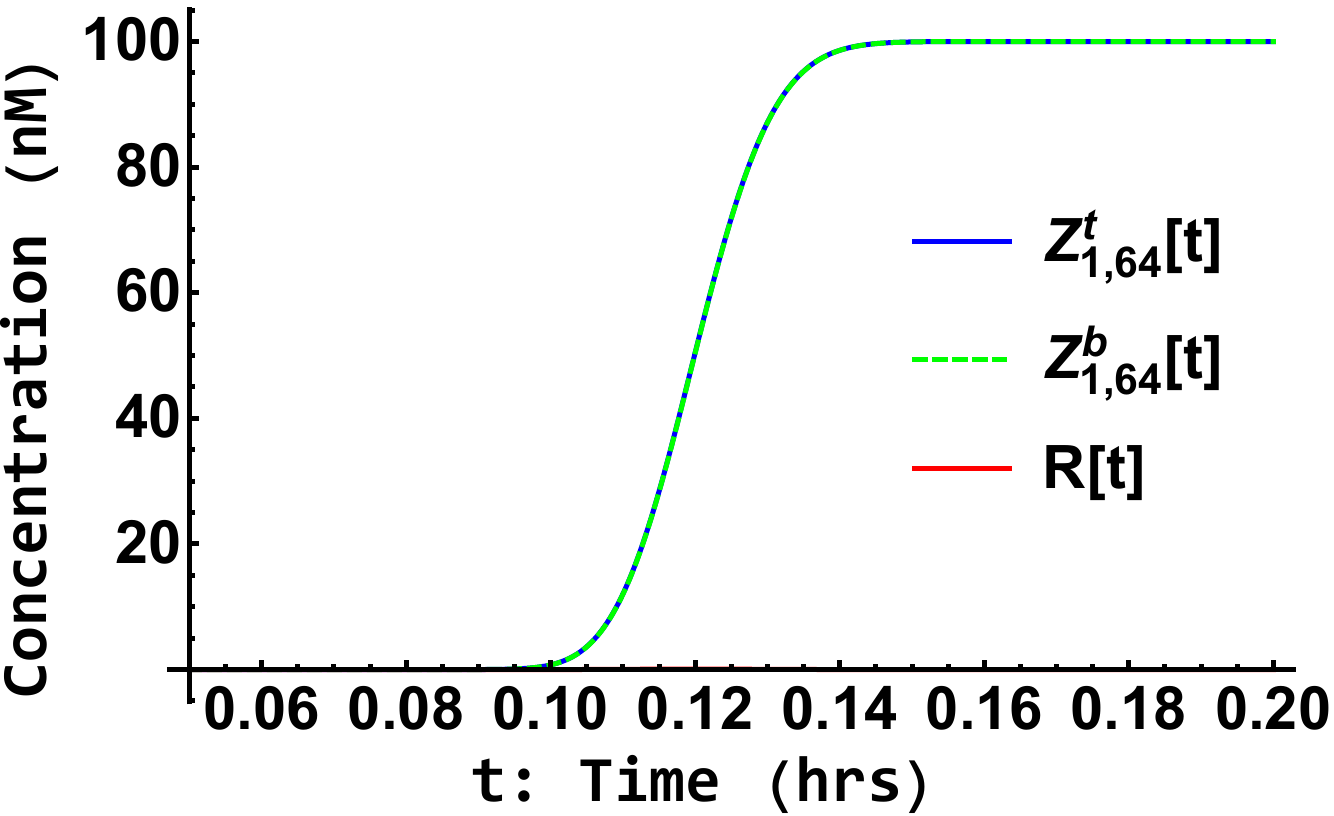}}
    \vspace{-4pt}
  \subfigure[\textcolor{black}{Enlarged version of Cond. 1 (R=1)}.]{
 \label{f3_14}
 \includegraphics[width=0.475\linewidth]{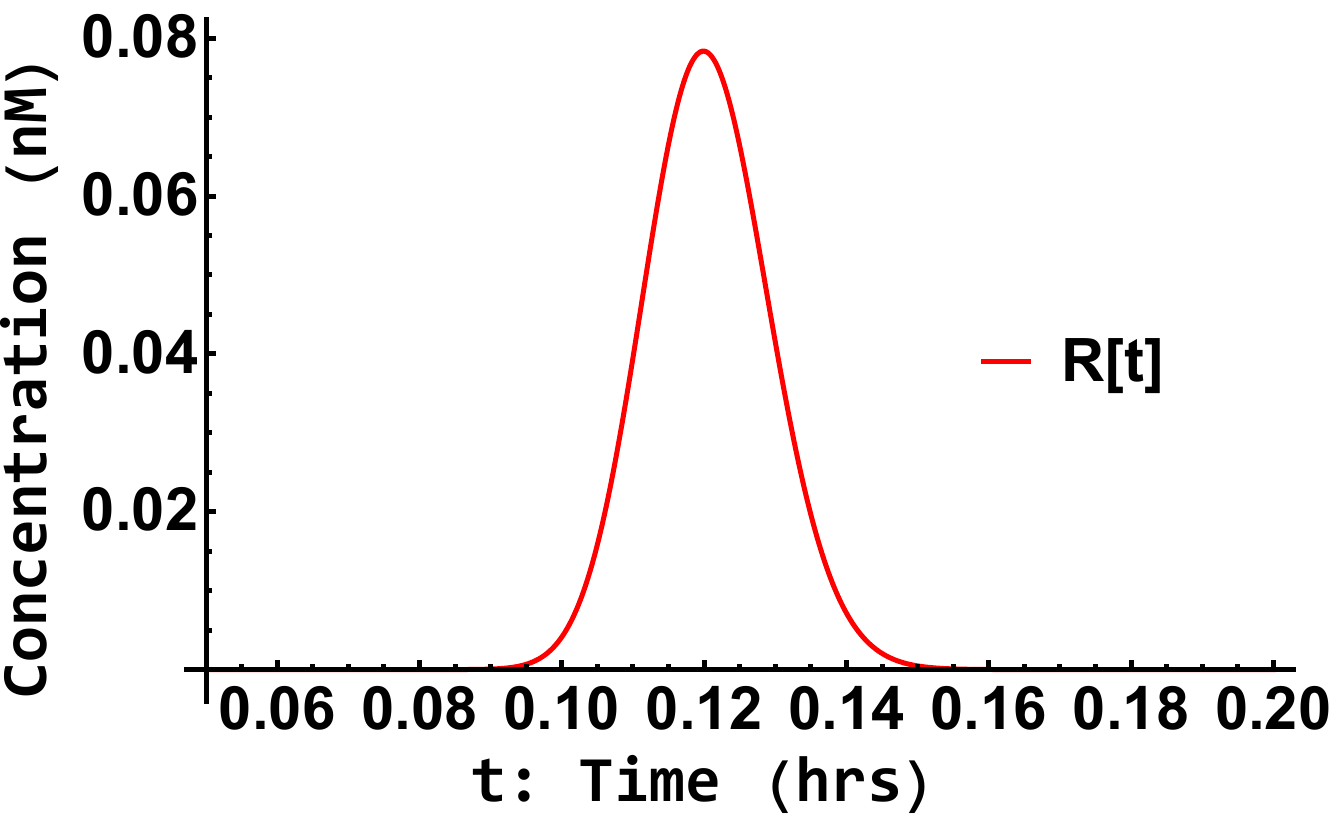}}
    \vspace{-4pt}
 \subfigure[ 64-stage PUF (Cond. 2) (R=0).]{
 \label{f3_15}
  \includegraphics[width=0.475\linewidth]{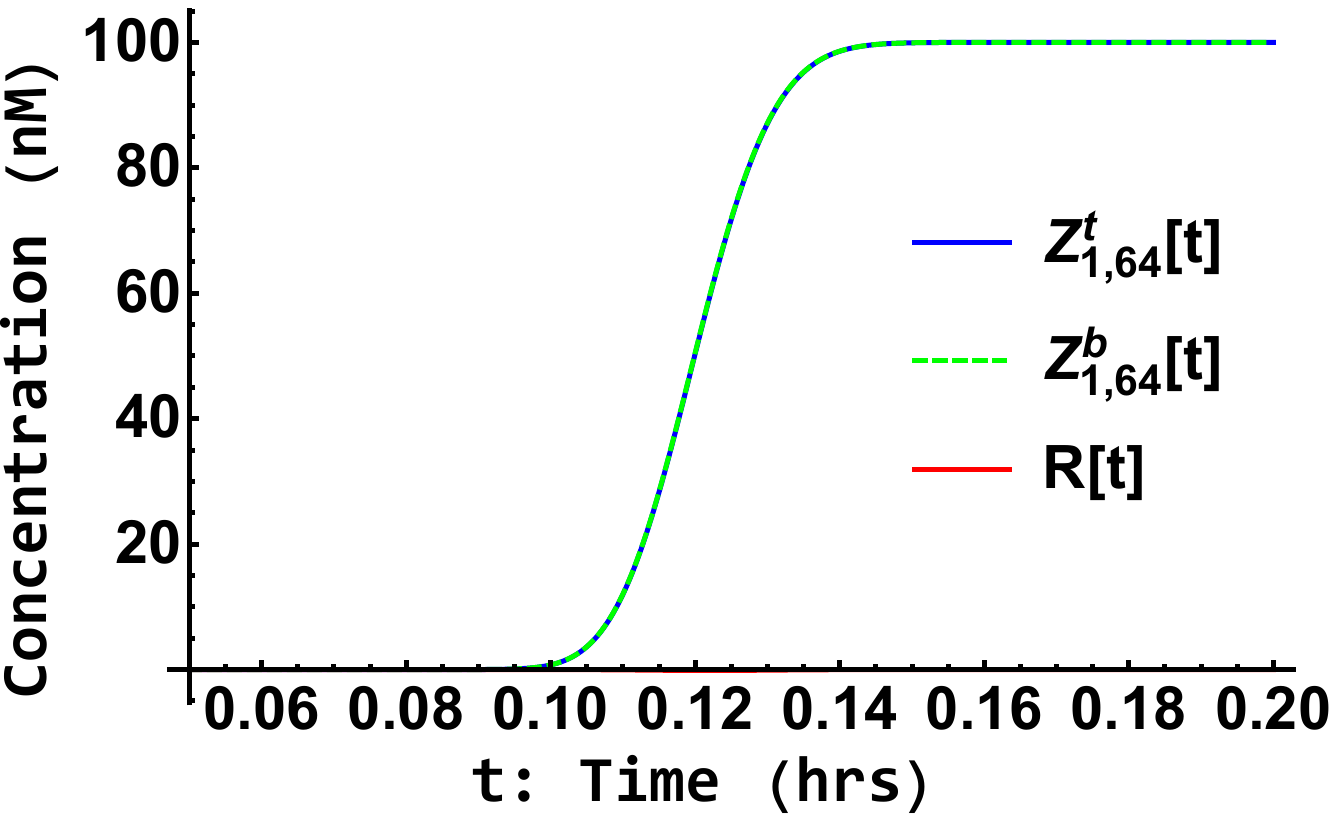}}
     \vspace{-4pt}
  \subfigure[Enlarged version of Cond. 2 (R=0).]{
 \label{f3_16}
  \includegraphics[width=0.475\linewidth]{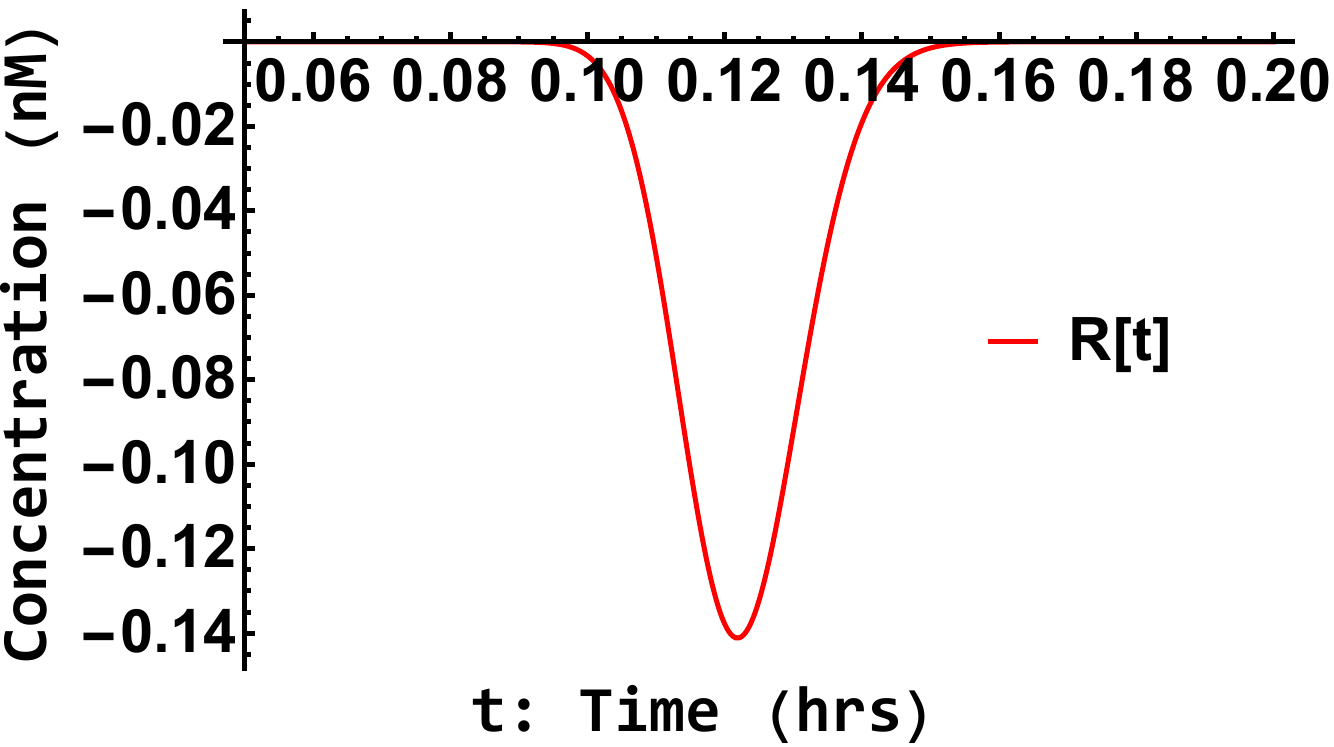}}
  \vspace{-2pt}
  \caption{Simulations for a single PUF (Continued).}%
\end{figure}
\vspace{-8pt}

\subsubsection{Inter-Chip Variation of Different PUFs Driven by the Same Challenge}
\begin{figure}[ht]
  \centering
    \subfigure[\textcolor{black}{8-stage PUF 1}.]{
 \label{f4_1}
 \includegraphics[width=0.475\linewidth]{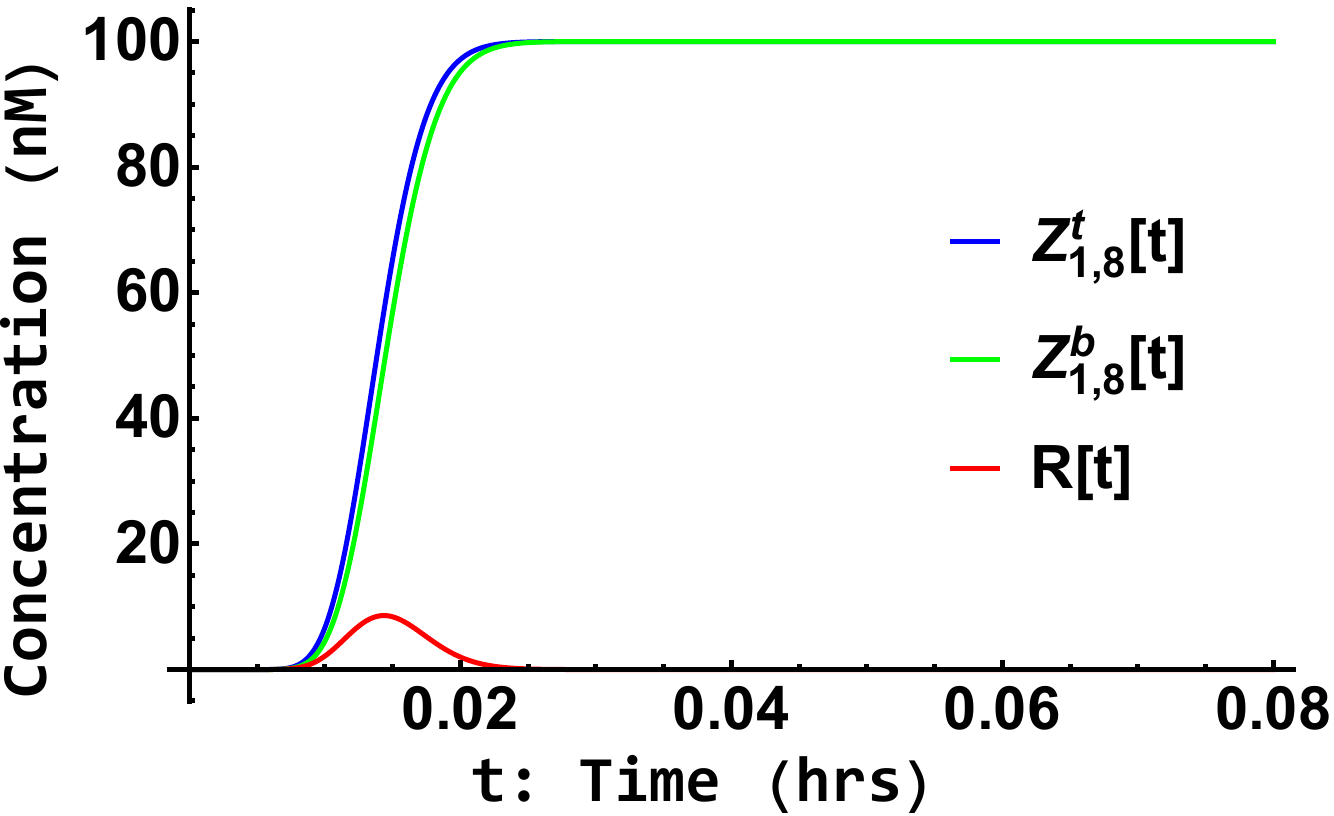}}
     \subfigure[\textcolor{black}{8-stage PUF 2}.]{
 \label{f4_5}
 \includegraphics[width=0.475\linewidth]{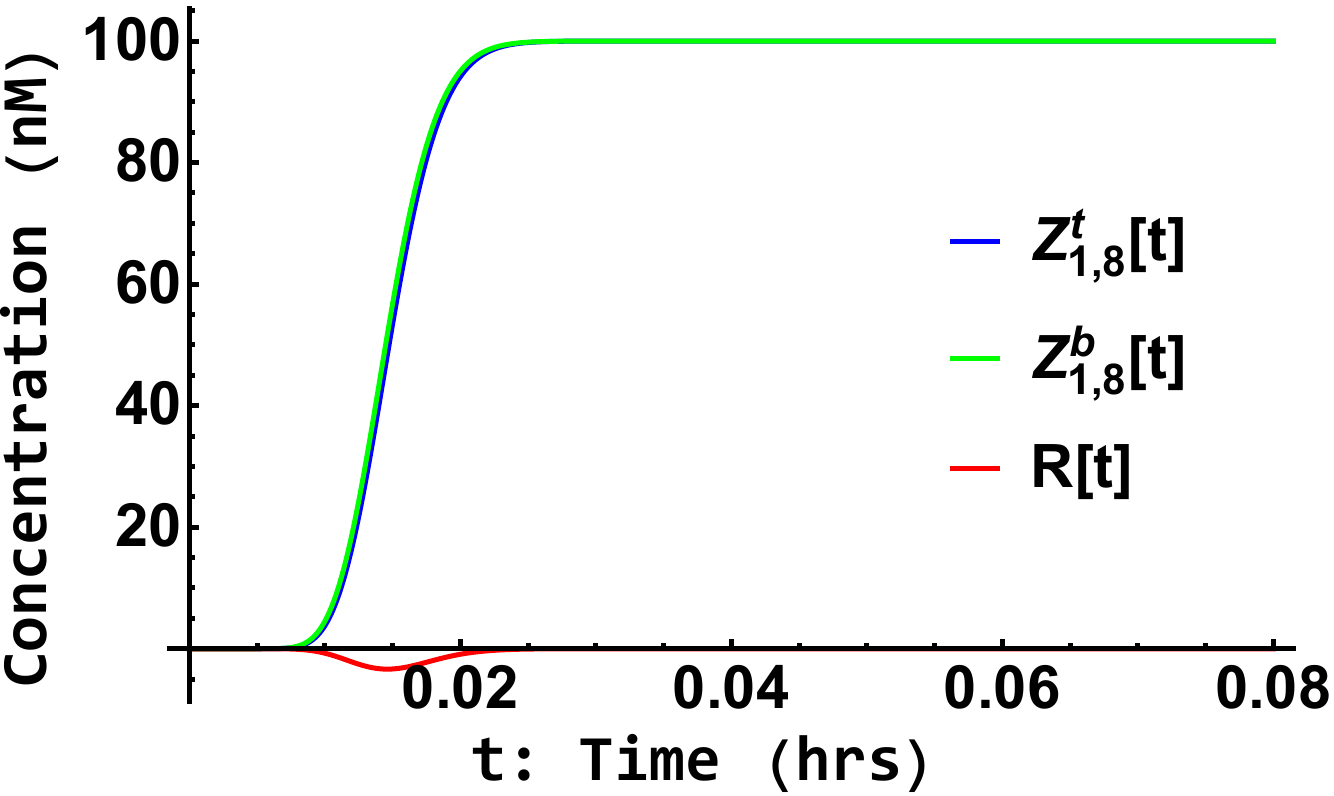}} \subfigure[\textcolor{black}{16-stage PUF 1}.]{
 \label{f4_2}
 \includegraphics[width=0.475\linewidth]{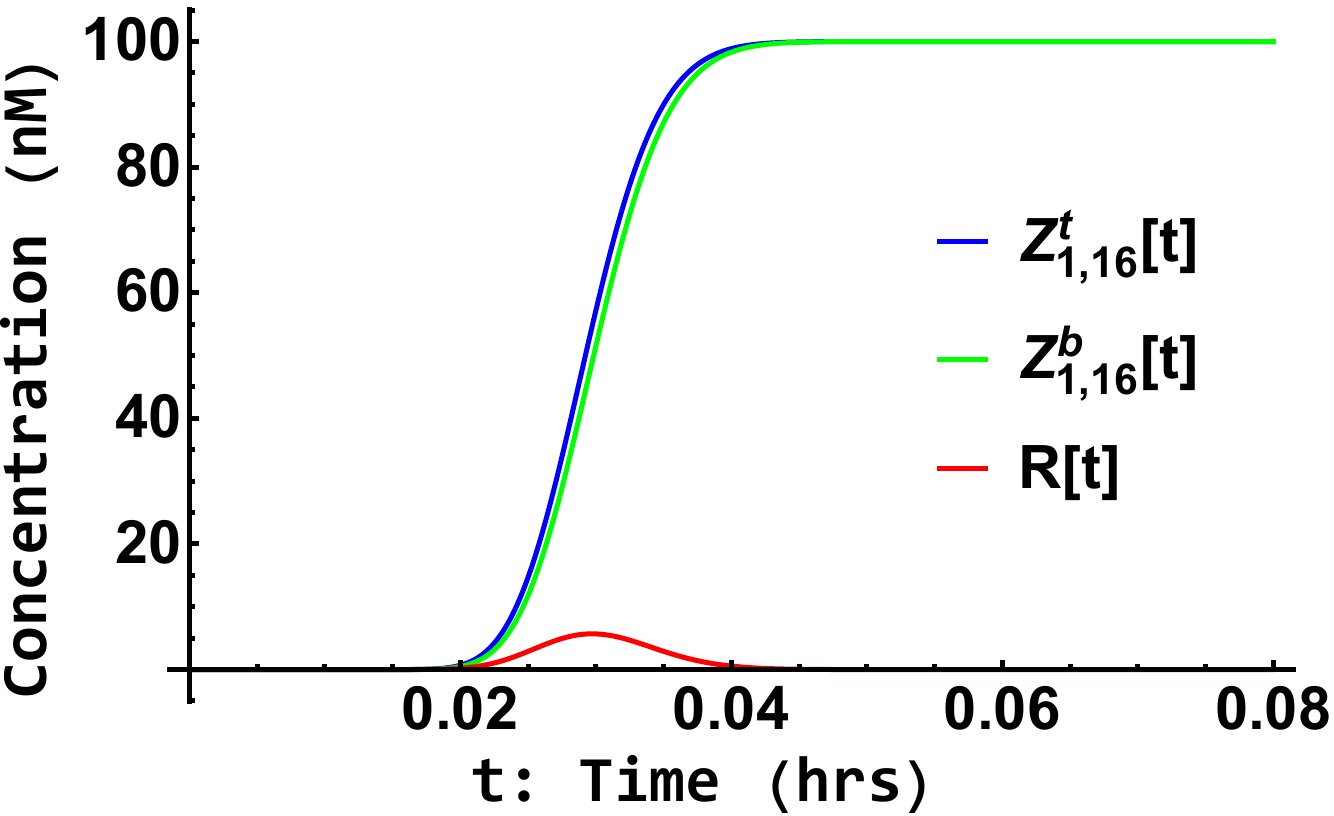}}
 \subfigure[\textcolor{black}{16-stage PUF 2}.]{
 \label{f4_6}
 \includegraphics[width=0.475\linewidth]{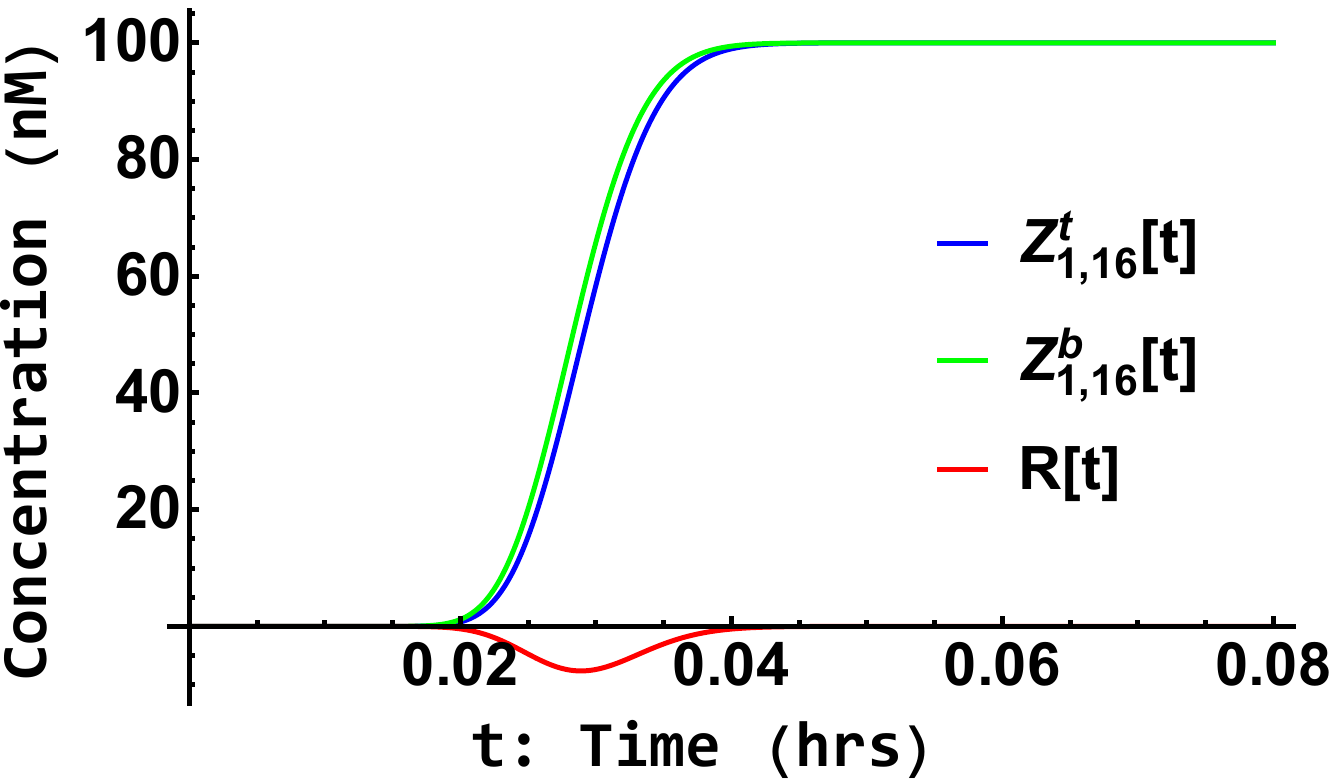}}
 \subfigure[32-stage PUF 1.]{
 \label{f4_3}
  \includegraphics[width=0.475\linewidth]{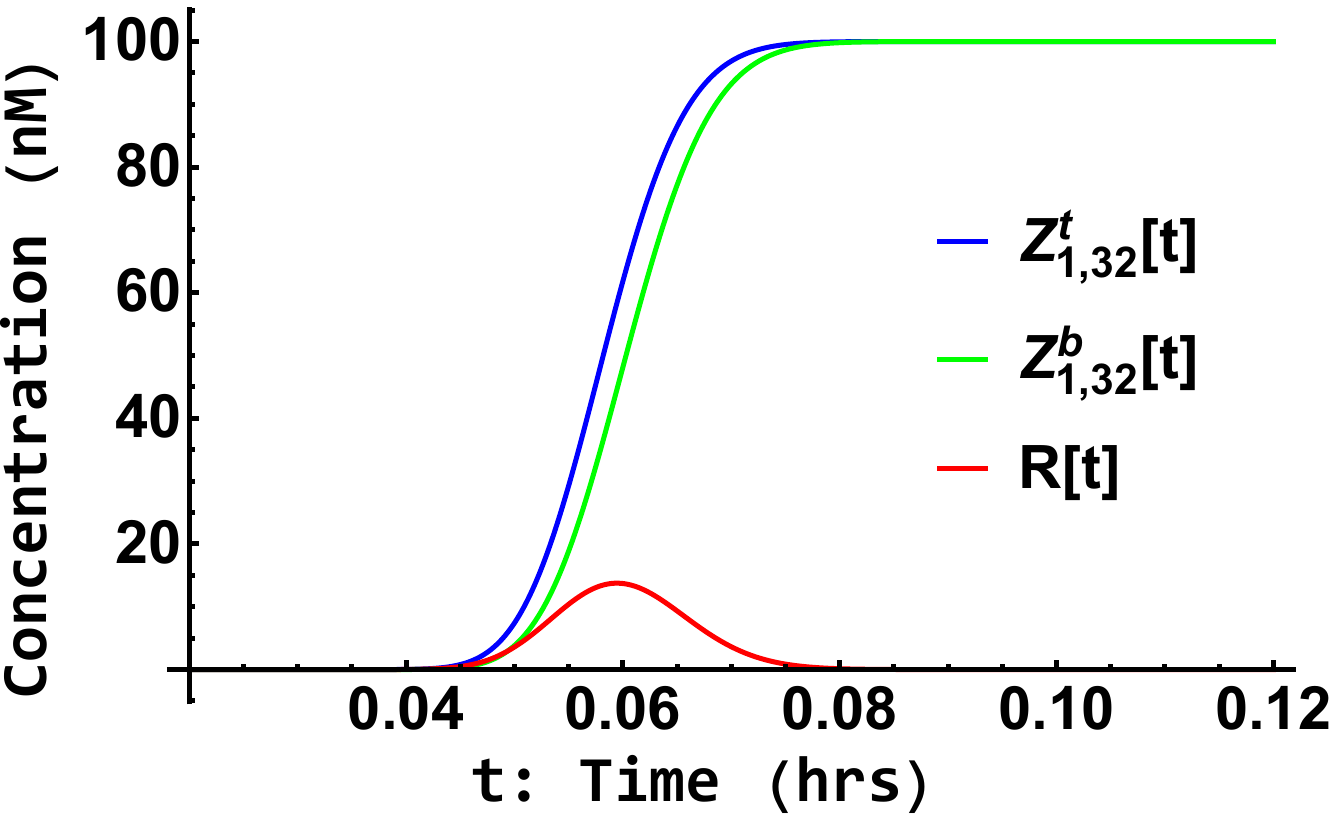}}
\subfigure[32-stage PUF 2.]{
 \label{f4_7}
  \includegraphics[width=0.475\linewidth]{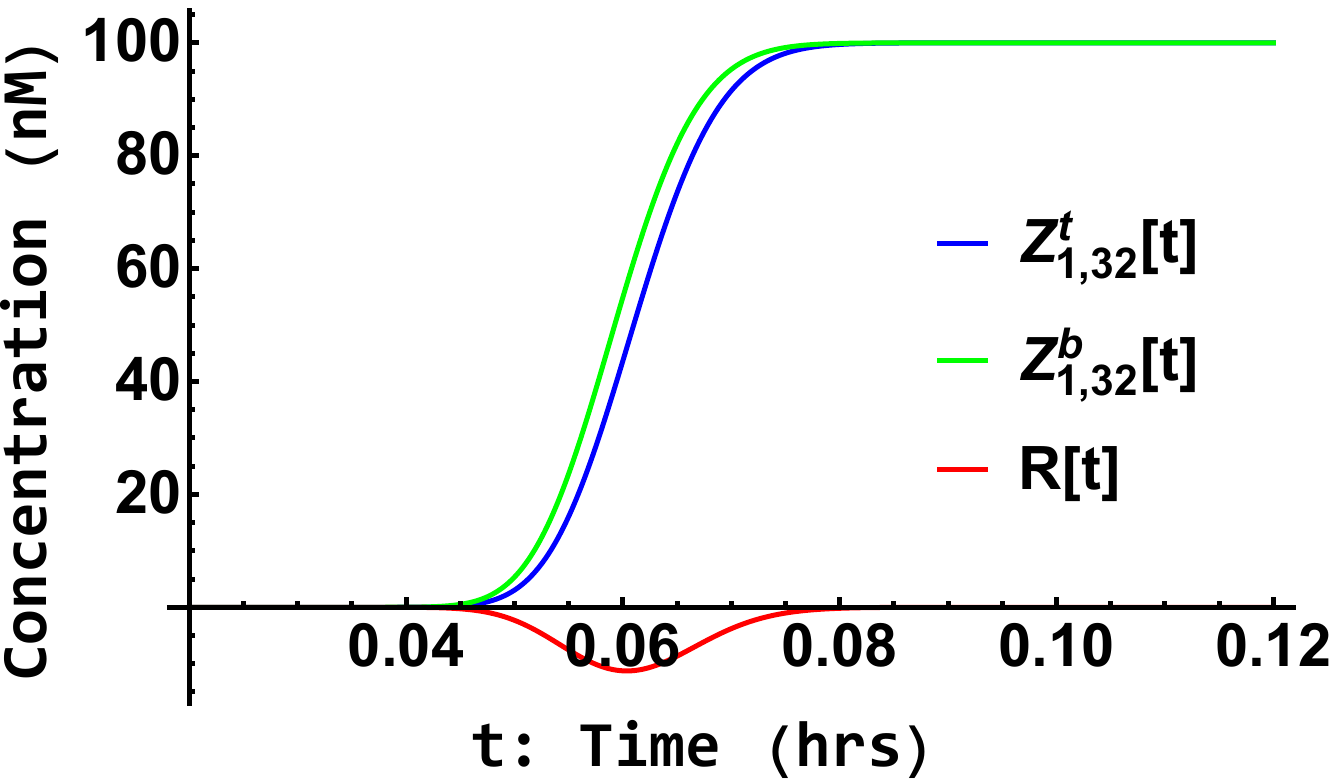}}   
  \subfigure[64-stage PUF 1.]{
 \label{f4_4}
  \includegraphics[width=0.475\linewidth]{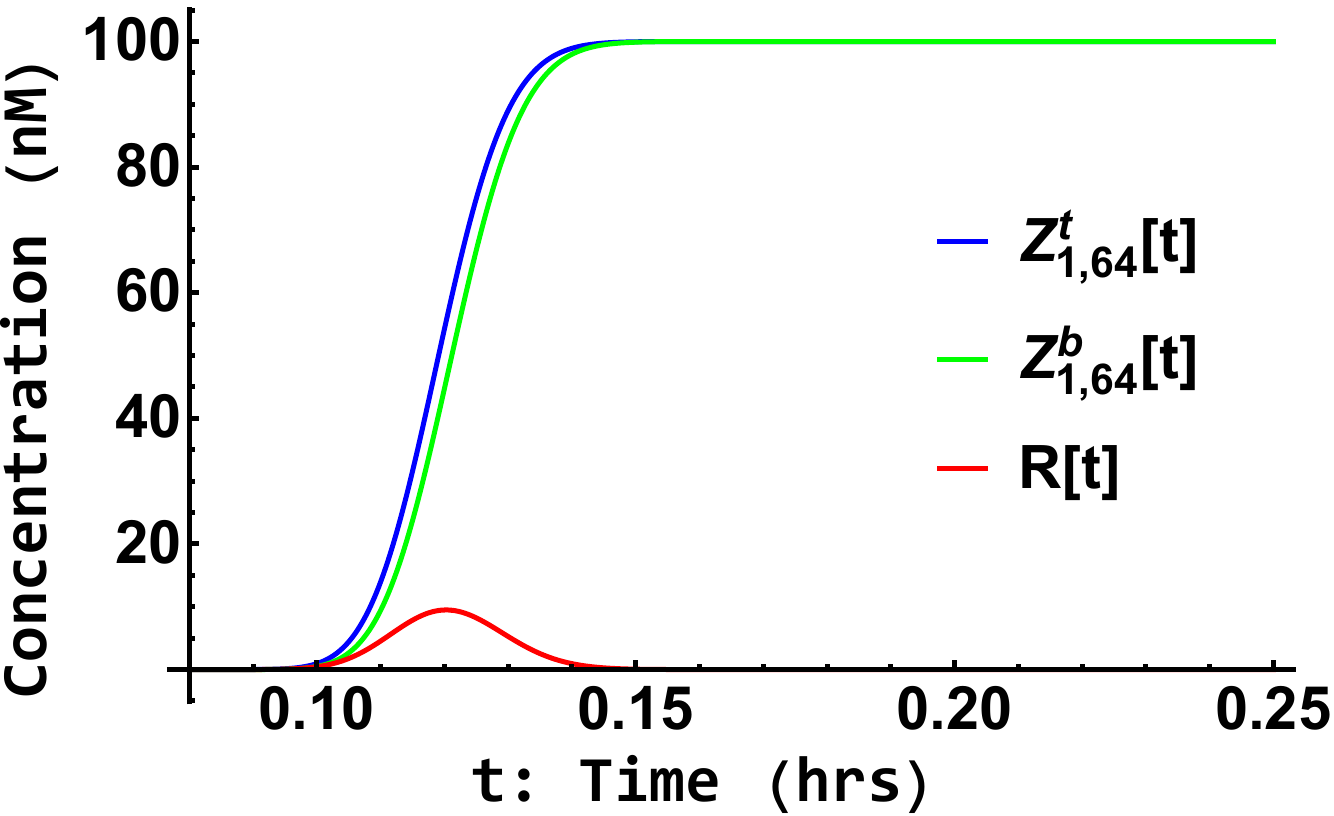}}
   \subfigure[64-stage PUF 2.]{
 \label{f4_8}
  \includegraphics[width=0.475\linewidth]{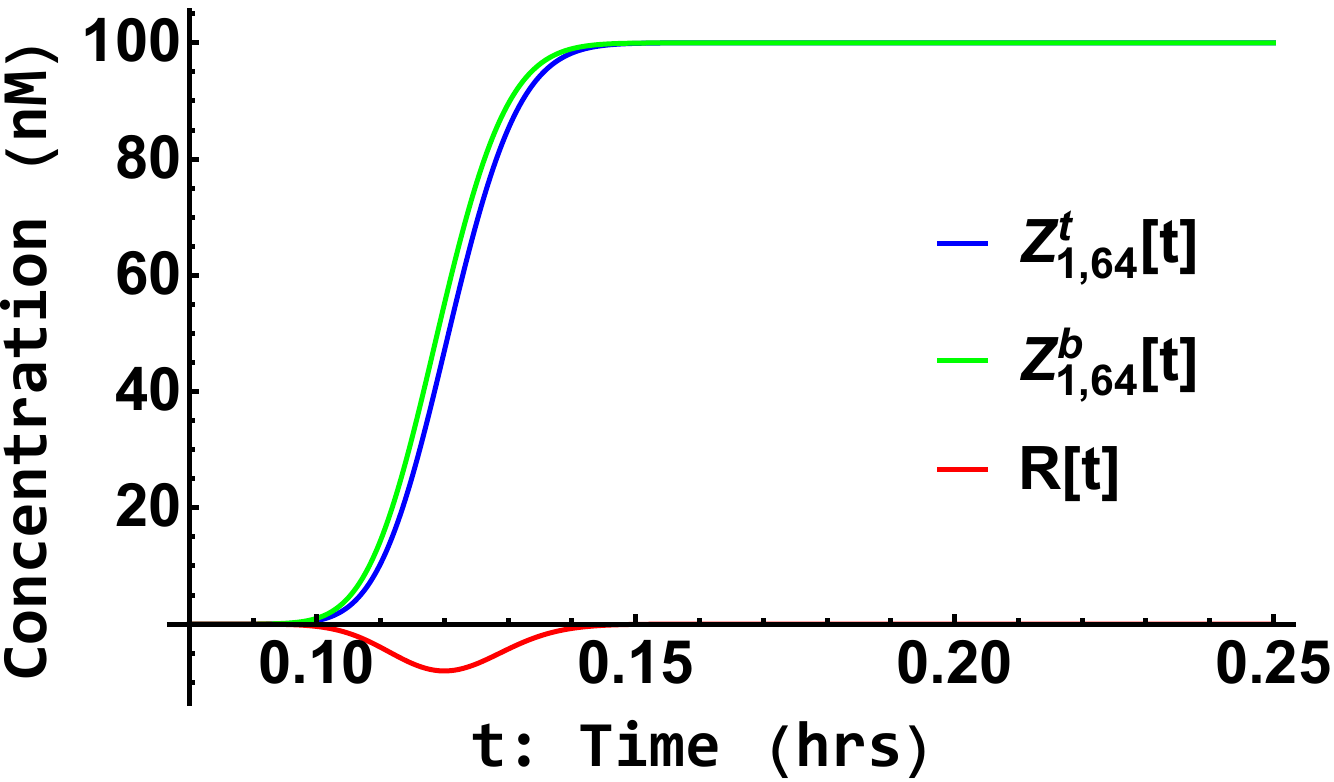}}
  \vspace{-7pt}
 \caption{Simulation results for $8$-stage, $16$-stage, $32$-stage and $64$-stage PUFs activated by the same challenge. Each challenge is applied to two different PUFs.}\label{f4}
\end{figure}
\vspace{-1pt}

\begin{table*}
  \centering
  \caption{Rate Constants for Top and Bottom MUXes for Two 8-Stage PUFs that Generate Different Responses\textcolor{red}{$^*$}}
  \vspace{-8pt}
    \begin{tabular}{c||c|c|c|c|c|c|c|c}
   \hline
   \multicolumn{9}{c}{\textbf{Specific Rate Constants for 8-stage PUF 1 to Generate the Positive Response}}\\
   \hline
    \textbf{Stage} & \textbf{1} & \textbf{2} & \textbf{3} & \textbf{4} & \textbf{5} & \textbf{6} & \textbf{7} & \textbf{8} \\
    \hline
    \hline
    \textbf{Top} & 16.2000  & 16.3874 & 17.8391 & 14.2102 & 15.8676 & 16.1017 & 15.4241 & 17.0567 \\
    \hline
    \textbf{Bottom} & 13.4213 & 17.7129 & 14.0757 & 16.2111 & 16.8199 & 14.8264 & 16.3124 & 15.0728 \\
    \hline
    \hline
%
   \multicolumn{9}{c}{\textbf{Specific Rate Constants for 8-stage PUF 2 to Generate the Negative Response}}\\
   \hline
    \textbf{Stage} & \textbf{1} & \textbf{2} & \textbf{3} & \textbf{4} & \textbf{5} & \textbf{6} & \textbf{7} & \textbf{8} \\
     \hline
     \hline
    \textbf{Top} & 15.3259 & 17.3474 & 15.3472 & 17.276 & 17.4534 & 15.7802 & 15.3876 & 16.8865 \\
    \hline
    \textbf{Bottom} & 16.0853 & 16.3489 & 16.2476 & 16.5559 & 15.937 & 14.9865 & 16.7846 & 16.8834 \\
    \hline
    \multicolumn{9}{l}{\textcolor{red}{$^*$}The $8$-bit challenge=[11101010].}
    \end{tabular}\label{tl}%
\end{table*}%
Fig. \ref{f4} shows four cases where two different PUFs of the same stage produce different responses, positive and negative, when driven by the same challenge under the same environmental condition. The rate constant schemes for all PUFs follow the Gaussian distribution $\mathcal{N}(16,1)$. Due to the page limit, rate constants for only two cases are listed in Table \ref{tl} for 8-stage PUFs and Table \ref{t2} for 16-stage PUFs, respectively. 

\begin{table*}[htbp]
  \centering
  \caption{Rate Constants for Top and Bottom MUXes for Two 16-Stage PUFs that Generate Different Responses\textcolor{red}{$^*$}}
  \vspace{-8pt}
    \begin{tabular}{c||c|c|c|c|c|c|c|c}
   \hline
   \multicolumn{9}{c}{\textbf{Specific Rate Constants for 16-stage PUF 1 to Generate the Positive Response}}\\
   \hline
    \textbf{Stage} & \textbf{1} & \textbf{2} & \textbf{3} & \textbf{4} & \textbf{5} & \textbf{6} & \textbf{7} & \textbf{8} \\
    \hline
     \hline
   \textbf{Top} &  16.2713 & 16.1154 & 16.0443 & 15.6363 & 14.8986 & 14.9884 & 14.2514 & 15.3905 \\
   \hline
   \textbf{Bottom}  & 15.7969 & 16.1660 & 16.1474 & 15.6422 & 17.0791 & 15.3631 & 15.5394 & 17.7289 \\
    \hline
     \hline
    \textbf{Stage} & \textbf{9} & \textbf{10} & \textbf{11} & \textbf{12} & \textbf{13} & \textbf{14} & \textbf{15} & \textbf{16} \\
    \hline
     \hline
  \textbf{Top} &  15.8301 & 15.8637 & 17.1239 & 17.0826 & 17.0086 & 16.9753 & 15.9769 & 15.2015 \\
  \hline
   \textbf{Bottom} & 14.0328 & 16.9466 & 14.6572 & 16.1119 & 15.6691 & 16.5813 & 16.5878 & 16.2887 \\
    \hline
   \multicolumn{9}{c}{\textbf{Specific Rate Constants for 16-stage PUF 2 to Generate the Negative Response}}\\
   \hline
    \textbf{Stage} & \textbf{1} & \textbf{2} & \textbf{3} & \textbf{4} & \textbf{5} & \textbf{6} & \textbf{7} & \textbf{8} \\
    \hline
     \hline
    \textbf{Top} & 16.0475 & 17.2968 & 15.6063 & 16.7512 & 16.7523 & 16.4987 & 16.9658 & 16.9509 \\
    \hline
    \textbf{Bottom} & 16.5777 & 16.2362 & 16.1138 & 14.3944 & 17.3143 & 16.2662 & 16.9262 & 14.7676 \\
    \hline
     \hline
    \textbf{Stage} & \textbf{9} & \textbf{10} & \textbf{11} & \textbf{12} & \textbf{13} & \textbf{14} & \textbf{15} & \textbf{16} \\
     \hline
    \hline
    \textbf{Top} & 17.6698 & 17.044 & 15.8186 & 15.9833 & 17.7202 & 17.3745 & 15.2481 & 16.2475 \\
    \hline
    \textbf{Bottom} & 16.1303 & 16.5323 & 16.0368 & 15.2344 & 16.5067 & 18.0267 & 16.7487 & 15.8938 \\
    \hline
    \multicolumn{9}{l}{\textcolor{red}{$^*$}The $16$-bit challenge=[0100000000110001].}
    \end{tabular}\label{t2}%
\end{table*}%

\begin{table*}[htbp]
  \centering
  \caption{Intra-Chip and Inter-Chip Variation Results for PUFs of 8, 16, 32 and 64 Stages}
  \vspace{-8pt}
    \begin{tabular}{c||ccc|ccc|ccc|ccc}
   \hline
    \textbf{Case} & \multicolumn{3}{c|}{\textbf{8-stage PUFs}} & \multicolumn{3}{c|}{\textbf{ 16-stage PUFs}} & \multicolumn{3}{c|}{\textbf{32-stage PUFs}} & \multicolumn{3}{c}{\textbf{ 64-stage PUFs}}\\
   \hline
    \textbf{Types} & \multicolumn{1}{c|}{\textbf{Max}} & \multicolumn{1}{c|}{\textbf{Min}} & \textbf{Mean}  & \multicolumn{1}{c|}{\textbf{Max}} & \multicolumn{1}{c|}{\textbf{Min}} & \textbf{Mean}  & \multicolumn{1}{c|}{\textbf{Max}} & \multicolumn{1}{c|}{\textbf{Min}} & \textbf{Mean} & \multicolumn{1}{c|}{\textbf{Max}}& \multicolumn{1}{c|}{\textbf{Min}} & \textbf{Mean}\\
   \hline
    \textbf{Intra-chip} & \multicolumn{1}{c|}{3.00\%} & \multicolumn{1}{c|}{0.00\%} & 0.24\% & \multicolumn{1}{c|}{1.00\%} & \multicolumn{1}{c|}{0.00\%} & 0.07\% & \multicolumn{1}{c|}{1.00\%} & \multicolumn{1}{c|}{0.00\%} & 0.37\% & \multicolumn{1}{c|}{ 1.5\%} & \multicolumn{1}{c|}{ 0.00\%} &  0.68\%\\
   \hline
    \textbf{Inter-chip} & \multicolumn{1}{c|}{95.0\%} & \multicolumn{1}{c|}{2.50\%} & 49.88\% & \multicolumn{1}{c|}{82.5\%} & \multicolumn{1}{c|}{15.00\%} & 49.98\% & \multicolumn{1}{c|}{78.5\%} & \multicolumn{1}{c|}{24.00\%} & 49.99\% & \multicolumn{1}{c|}{ 68.5\%} & \multicolumn{1}{c|}{ 30.00\%} &  50.00\%\\
   \hline
   $Reliability$ & \multicolumn{3}{c|}{99.76\%} & \multicolumn{3}{c|}{99.93\%} & \multicolumn{3}{c|}{99.63\%} &\multicolumn{3}{c}{ 99.32\%}\\
   \hline
   $ Uniqueness$ & \multicolumn{3}{c|}{99.76\%} & \multicolumn{3}{c|}{99.95\%} & \multicolumn{3}{c|}{99.97\%}&\multicolumn{3}{c}{ 100\%} \\
   \hline
    \end{tabular}\label{t3}%
\end{table*}%

\begin{figure*}[htbp]
  \centering
    \subfigure[\textcolor{black}{8-stage PUFs}.]{
 \label{f5_1}
 \includegraphics[width=0.235\linewidth]{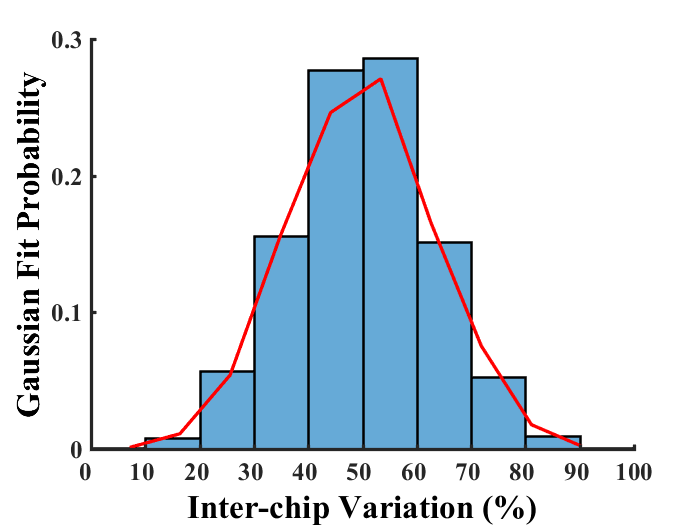}}
  \subfigure[\textcolor{black}{16-stage PUFs}.]{
 \label{f5_2}
 \includegraphics[width=0.235\linewidth]{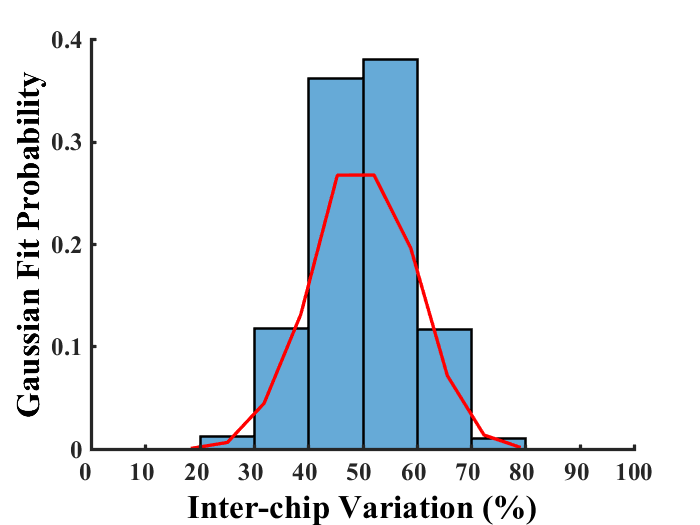}}
 \subfigure[32-stage PUFs.]{
 \label{f5_3}
  \includegraphics[width=0.235\linewidth]{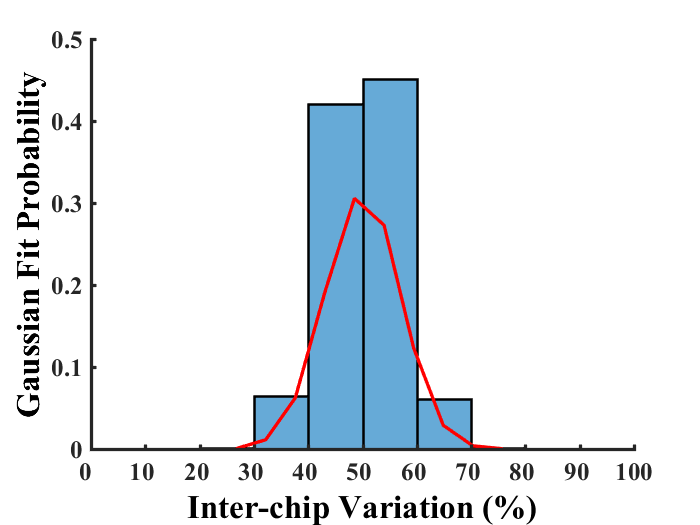}}
   \subfigure[ 64-stage PUFs.]{
 \label{f5_4}
  \includegraphics[width=0.235\linewidth]{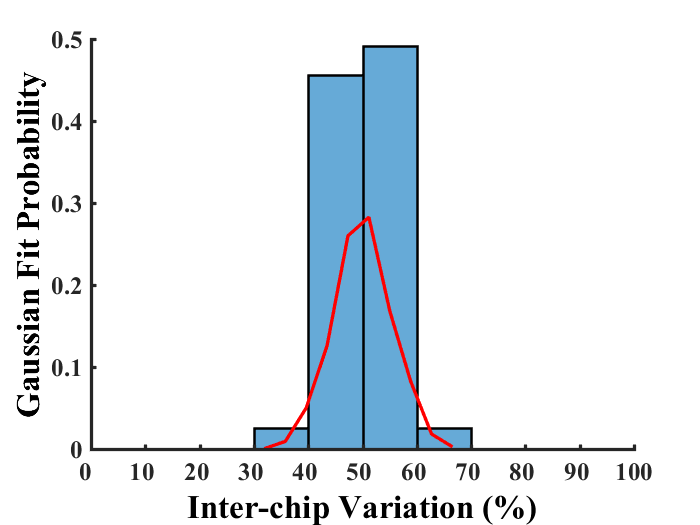}}
  \vspace{-6pt}
 \caption{Gaussian fit curve of inter-chip variation distribution for $8$-stage, $16$-stage, $32$-stage and $64$-stage PUFs.}\label{f5}
\end{figure*}

Notice that, as the number of stages increases, the arrival time at the $N$th-stage of both top and bottom inputs is prolonged. Fig. \ref{f4} shows that different PUFs can produce different responses when driven by the same challenge.


%



\subsection{PUF Performance Metrics}
The results in Section \ref{sec:puf} indicate that slight variations in rate constant of each MUX indeed provide sufficient randomness for the PUF. Thus for different PUFs, the responses can be varied even under the same challenge bits. This subsection calculates the aforementioned two metrics to analyze the synthesized PUFs' performance.

\subsubsection{Reliability} 
This metric is measured for a single $N$-stage PUF under $m$ environmental conditions with $N$-bit challenges to generate the response $R$ with the length of $L$ bits. The noise caused by environment follows the Gaussian distribution $\mathcal{N}(0, \sigma_s^2)$, where the standard deviation $\sigma_s=0.05$ in the whole paper. To be more specific, we use $m=200$ environmental conditions to obtain the response $R$ with the bit length $L=200$. Thus, the number of possible piecewise comparison for \eqref{eq1:0} is $m-1=200-1=199$.

\subsubsection{Uniqueness}
This metric is evaluated by $K$ different $N$-stage PUFs where the rate constant is sampled from a Gaussian distribution $\mathcal{N}(16, 1)$. Specifically, we use $K=200$ different PUFs to generate the response $R$ with the bit length $L=200$. Therefore, for \eqref{eq2:0}, we have totally ${K \choose 2}={200 \choose 2}=19900$ piecewise comparisons.

Here we study four types of PUFs with the number of stages $8$, $16$, $32$ and $64$. The corresponding intra-chip and inter-chip variations are shown in Table \ref{t3}. Based on \eqref{eq1:0} and \eqref{eq2:0}, the calculated two PUF performance metrics, reliability and uniqueness, are also listed in this table. Fig. \ref{f5} shows the Gaussian fit curve of inter-chip variation distribution for different stages.

Prior to analysis, it should be emphasized that the minimum inter-chip variation should be in practice larger than the maximum intra-chip variation  \cite{scheffer2002explicit}. \textcolor{blue}{\textit{\textbf{1).}}} Due to this reason, the $8$-stage PUF in Table \ref{t3} is not feasible as a PUF. Thus, we only analyze the remaining three cases, $16$, $32$ and $64$-stage PUFs. \textcolor{blue}{\textit{\textbf{2).}}} Based on Fig. \ref{f5}, the shape of Gaussian fit curve gets narrower and the minimum inter-chip variation increases with the number of stages. This indicates that the number of stages has an obvious impact on inter-chip variation. \textcolor{blue}{\textit{\textbf{3).}}} Comparing the three PUFs, as the number of stages increases, the intra-chip variation also increases, but is still about $1 \%$, while the inter-chip variation is closer to $50\%$. \textcolor{blue}{\textit{\textbf{4).}}} With respect to the reliability metric, $16$-stage PUFs show the highest reliability, while $64$-stage PUFs have the least reliability. With respect to the uniqueness metric, $64$-stage PUFs achieve the highest uniqueness. \textcolor{blue}{\textit{\textbf{5).}}} From these cases, we note that larger the number of stages, the more likely the synthesized molecular PUF will satisfy that the maximum intra-chip variation is less than the minimum inter-chip variation. 


According to the derived results for silicon MUX-based PUFs in \cite{lao2014statistical}, we know that with increase in the number of stages, $N$, \textit{1).} $P_{intra}$ increases, and the reliability decreases. \textit{2).} $P_{inter}$ increases and the uniqueness increases. The results from the molecular PUFs are consistent with these observations.

\section{Complexity}\label{s:4}
In a general case, to synthesize an $N$-stage molecular PUF, as expressed in \eqref{eq4}, a total of $32N$ chemical reactions are required. Each MUX requires $16$ reactions and the PUF contains $2N$ multiplexers. 
\vspace{-8pt}
\begin{equation}\label{eq4}
\#_{\text{reactions}}=2N \times 16=32N
\end{equation}
 


\section{Conclusion}\label{s:5}
This paper has investigated the feasibility of a molecular MUX PUF using molecular reactions. The randomness in the rate constant is an assumption that has been used in this paper; its validity in an experimental setup remains to be demonstrated. Although the paper demonstrates feasibility of a molecular PUF, several limitations exist in the current implementation. For the molecular PUF to be complete, a molecular arbiter should be used to compute the response. One possible realization is the use of the D latch presented in \cite{jiang2013digital}. The need to initialize the top and bottom outputs of each MUX of each stage after each authentication is also a limitation of the proposed PUF. Molecular PUFs that do not suffer from this limitation should be investigated. The rate constant for MUX stages is sampled from the Gaussian distribution $\mathcal{N}(16, 1)$ in this paper. The performance of a molecular PUF is dependent on the variance of the rate constant. In practice the variability of the rate constant is dependent on the kinetics of strand displacement that can be modulated by parameters such as toeholds \cite{zhang2009control}. Successful experimental demonstration of a molecular PUF is a topic of future research. Similar to electronic PUFs \cite{koyily2018effect}, effects of aging, changes in heat, light and other environmental noise on molecular PUF need to be understood.

While the feasibility of a molecular MUX PUF is of interest, demonstrating feasibility of DNA PUFs is of greater interest. Thus, molecular reactions need to be mapped to DNA. Similar to molecular PUFs, if the number of stages is small, the DNA PUF may not be useful as a PUF structure. The minimum number of stages needed for a DNA PUF to be feasible needs to be investigated. 

In electronic PUFs, path delay differences for each stage of the MUX PUF can be learned by a software model such as an ANN \cite{ruhrmair2013puf,avvaru2016estimating}. Demonstrating the ability to learn the model of a molecular PUF by software models from experimental data needs to be investigated. In recent work, secure yet reliable electronic MUX PUFs have been demonstrated using homogeneous XOR PUFs \cite{zhou2017secure} and heterogeneous XOR PUFs \cite{avvaru2020homogeneous}. Investigating security and reliability of homogeneous and heterogeneous molecular XOR PUFs is a topic of future research.
\vspace{-4pt}
\section*{Acknowledgment}
The authors thank Xingyi Liu for numerous valuable discussions. L. Ge has been supported by the Chinese Scholarship Council (CSC). 
\vspace{-4pt}



\footnotesize
\bibliographystyle{IEEEtran}
\bibliography{IEEEabrv,mybib}

\begin{thebibliography}{10}
\providecommand{\url}[1]{#1}
\csname url@samestyle\endcsname
\providecommand{\newblock}{\relax}
\providecommand{\bibinfo}[2]{#2}
\providecommand{\BIBentrySTDinterwordspacing}{\spaceskip=0pt\relax}
\providecommand{\BIBentryALTinterwordstretchfactor}{4}
\providecommand{\BIBentryALTinterwordspacing}{\spaceskip=\fontdimen2\font plus
\BIBentryALTinterwordstretchfactor\fontdimen3\font minus
  \fontdimen4\font\relax}
\providecommand{\BIBforeignlanguage}[2]{{%
\expandafter\ifx\csname l@#1\endcsname\relax
\typeout{** WARNING: IEEEtran.bst: No hyphenation pattern has been}%
\typeout{** loaded for the language `#1'. Using the pattern for}%
\typeout{** the default language instead.}%
\else
\language=\csname l@#1\endcsname
\fi
#2}}
\providecommand{\BIBdecl}{\relax}
\BIBdecl

\bibitem{pappu2002physical}
R.~Pappu, B.~Recht, J.~Taylor, and N.~Gershenfeld, ``Physical one-way
  functions,'' \emph{Science}, vol. 297, no. 5589, pp. 2026--2030, 2002.

\bibitem{gassend2002silicon}
B.~Gassend, D.~Clarke, M.~Van~Dijk, and S.~Devadas, ``Silicon physical random
  functions,'' in \emph{Proceedings of the 9th ACM conference on Computer and
  communications security}.\hskip 1em plus 0.5em minus 0.4em\relax ACM, 2002,
  pp. 148--160.

\bibitem{suh2007physical}
G.~E. Suh and S.~Devadas, ``Physical unclonable functions for device
  authentication and secret key generation,'' in \emph{2007 44th ACM/IEEE
  Design Automation Conference}.\hskip 1em plus 0.5em minus 0.4em\relax IEEE,
  2007, pp. 9--14.

\bibitem{lao2014statistical}
Y.~Lao and K.~K. Parhi, ``Statistical analysis of {MUX}-based physical
  unclonable functions,'' \emph{IEEE Trans. Comput.-Aided Des. Integr. Circuits
  Syst.}, vol.~33, no.~5, pp. 649--662, 2014.

\bibitem{gao2016emerging}
Y.~Gao, D.~C. Ranasinghe, S.~F. Al-Sarawi, O.~Kavehei, and D.~Abbott,
  ``Emerging physical unclonable functions with nanotechnology,'' \emph{IEEE
  access}, vol.~4, pp. 61--80, 2016.

\bibitem{arppe2017physical}
R.~Arppe and T.~J. S{\o}rensen, ``Physical unclonable functions generated
  through chemical methods for anti-counterfeiting,'' \emph{Nature Reviews
  Chemistry}, vol.~1, no.~4, p. 0031, 2017.

\bibitem{armknecht2011formalization}
F.~Armknecht, R.~Maes, A.-R. Sadeghi, F.-X. Standaert, and C.~Wachsmann, ``A
  formalization of the security features of physical functions,'' in \emph{2011
  IEEE Symposium on Security and Privacy}.\hskip 1em plus 0.5em minus
  0.4em\relax IEEE, 2011, pp. 397--412.

\bibitem{chatterjee2018rf}
B.~Chatterjee, D.~Das, S.~Maity, and S.~Sen, ``{RF-PUF}: Enhancing {IoT}
  security through authentication of wireless nodes using in-situ machine
  learning,'' \emph{IEEE Internet of Things Journal}, vol.~6, no.~1, pp.
  388--398, 2018.

\bibitem{johnson2016verifying}
R.~F. Johnson, Q.~Dong, and E.~Winfree, ``Verifying chemical reaction network
  implementations: a bisimulation approach,'' in \emph{International Conference
  on DNA-Based Computers}.\hskip 1em plus 0.5em minus 0.4em\relax Springer,
  2016, pp. 114--134.

\bibitem{cardelli2008processes}
L.~Cardelli, ``From processes to {ODEs} by chemistry,'' in \emph{Fifth IFIP
  International Conference On Theoretical Computer Science--Tcs 2008}.\hskip
  1em plus 0.5em minus 0.4em\relax Springer, 2008, pp. 261--281.

\bibitem{soloveichik2010dna}
D.~Soloveichik, G.~Seelig, and E.~Winfree, ``{DNA} as a universal substrate for
  chemical kinetics,'' \emph{Proc. Natl. Acad. Sci. U.S.A}, vol. 107, no.~12,
  pp. 5393--5398, 2010.

\bibitem{chen2013programmable}
Y.-J. Chen, N.~Dalchau, N.~Srinivas, A.~Phillips, L.~Cardelli, D.~Soloveichik,
  and G.~Seelig, ``Programmable chemical controllers made from {DNA},''
  \emph{Nat. Nanotechnol.}, vol.~8, no.~10, p. 755, 2013.

\bibitem{ge2017formal}
L.~Ge, Z.~Zhong, D.~Wen, X.~You, and C.~Zhang, ``A formal combinational logic
  synthesis with chemical reaction networks,'' \emph{IEEE Trans. Mol. Biol.
  Multi-Scale Commun.}, vol.~3, no.~1, pp. 33--47, 2017.

\bibitem{jiang2013digital}
H.~Jiang, M.~D. Riedel, and K.~K. Parhi, ``Digital logic with molecular
  reactions,'' in \emph{2013 IEEE/ACM International Conference on
  Computer-Aided Design (ICCAD)}.\hskip 1em plus 0.5em minus 0.4em\relax IEEE,
  2013, pp. 721--727.

\bibitem{jiang2011synchronous}
H.~Jiang, M.~Riedel, and K.~Parhi, ``Synchronous sequential computation with
  molecular reactions,'' in \emph{Proceedings of the 48th Design Automation
  Conference}.\hskip 1em plus 0.5em minus 0.4em\relax ACM, 2011, pp. 836--841.

\bibitem{salehi2015molecular}
S.~A. Salehi, H.~Jiang, M.~D. Riedel, and K.~K. Parhi, ``Molecular sensing and
  computing systems,'' \emph{IEEE Transactions on Molecular, Biological and
  Multi-Scale Communications}, vol.~1, no.~3, pp. 249--264, 2015.

\bibitem{kharam2011binary}
A.~Kharam, H.~Jiang, M.~Riedel, and K.~Parhi, ``Binary counting with chemical
  reactions,'' in \emph{Biocomputing 2011}.\hskip 1em plus 0.5em minus
  0.4em\relax World Scientific, 2011, pp. 302--313.

\bibitem{holden2019encrypted}
M.~T. Holden and L.~M. Smith, ``Encrypted oligonucleotide arrays for molecular
  authentication,'' \emph{ACS combinatorial science}, vol.~21, no.~8, pp.
  562--567, 2019.

\bibitem{shokralla2015dna}
S.~Shokralla, R.~S. Hellberg, S.~M. Handy, I.~King, and M.~Hajibabaei, ``A
  {DNA} mini-barcoding system for authentication of processed fish products,''
  \emph{Scientific Reports}, vol.~5, p. 15894, 2015.

\bibitem{qian2011scaling}
L.~Qian and E.~Winfree, ``Scaling up digital circuit computation with {DNA}
  strand displacement cascades,'' \emph{Science}, vol. 332, no. 6034, pp.
  1196--1201, 2011.

\bibitem{maiti2013systematic}
A.~Maiti, V.~Gunreddy, and P.~Schaumont, ``A systematic method to evaluate and
  compare the performance of physical unclonable functions,'' in \emph{Embedded
  systems design with FPGAs}.\hskip 1em plus 0.5em minus 0.4em\relax Springer,
  2013, pp. 245--267.

\bibitem{scheffer2002explicit}
L.~Scheffer, ``Explicit computation of performance as a function of process
  variation,'' in \emph{Proceedings of the 8th ACM/IEEE international workshop
  on Timing issues in the specification and synthesis of digital
  systems}.\hskip 1em plus 0.5em minus 0.4em\relax ACM, 2002, pp. 1--8.

\bibitem{zhang2009control}
D.~Y. Zhang and E.~Winfree, ``Control of {DNA} strand displacement kinetics
  using toehold exchange,'' \emph{Journal of the American Chemical Society},
  vol. 131, no.~47, pp. 17\,303--17\,314, 2009.

\bibitem{koyily2018effect}
A.~Koyily, S.~V.~S. Avvaru, C.~Zhou, C.~H. Kim, and K.~K. Parhi, ``Effect of
  aging on linear and nonlinear {MUX PUFs} by statistical modeling,'' in
  \emph{2018 23rd Asia and South Pacific Design Automation Conference
  (ASP-DAC)}.\hskip 1em plus 0.5em minus 0.4em\relax IEEE, 2018, pp. 76--83.

\bibitem{ruhrmair2013puf}
U.~R{\"u}hrmair, J.~S{\"o}lter, F.~Sehnke, X.~Xu, A.~Mahmoud, V.~Stoyanova,
  G.~Dror, J.~Schmidhuber, W.~Burleson, and S.~Devadas, ``{PUF} modeling
  attacks on simulated and silicon data,'' \emph{IEEE transactions on
  information forensics and security}, vol.~8, no.~11, pp. 1876--1891, 2013.

\bibitem{avvaru2016estimating}
S.~S. Avvaru, C.~Zhou, S.~Satapathy, Y.~Lao, C.~H. Kim, and K.~K. Parhi,
  ``Estimating delay differences of arbiter {PUFs} using silicon data,'' in
  \emph{2016 Design, Automation \& Test in Europe Conference \& Exhibition
  (DATE)}.\hskip 1em plus 0.5em minus 0.4em\relax IEEE, 2016, pp. 543--546.

\bibitem{zhou2017secure}
C.~Zhou, K.~K. Parhi, and C.~H. Kim, ``Secure and reliable {XOR} arbiter {PUF}
  design: An experimental study based on 1 trillion challenge response pair
  measurements,'' in \emph{Proceedings of the 54th Annual Design Automation
  Conference 2017}, 2017, pp. 1--6.

\bibitem{avvaru2020homogeneous}
S.~S. Avvaru, Z.~Zeng, and K.~K. Parhi, ``Homogeneous and heterogeneous
  feed-forward {XOR} physical unclonable functions,'' \emph{IEEE Transactions
  on Information Forensics and Security}, vol.~15, pp. 2485--2498, 2020.

\end{thebibliography}

\end{document}